\documentclass[11pt, a4paper]{article}
\usepackage{latexsym}
\usepackage{amsfonts}
\usepackage{pxfonts}
\usepackage[dvips]{graphics,epsfig}
\usepackage{subfigure}

\font\tenbb=msbm10
\font\sevenbb=msbm7
\font\fivebb=msbm5
\newfam\bbfam
\textfont\bbfam=\tenbb
\scriptfont\bbfam=\sevenbb
\scriptscriptfont\bbfam=\fivebb

\def\cqfd{$\sqcup\!\!\!\!\sqcap$}

\newcounter{Rem}
\refstepcounter{Rem}

\begin{document}
\vskip8ex
\setlength{\abovedisplayskip}{0.25cm}
\setlength{\belowdisplayskip}{0.25cm}

\newtheorem{Remark}{Remark}
\newtheorem{theo}{Theorem}
\newtheorem{prop}{Proposition}
\newtheorem{defi}{Definition}
\newtheorem{lem}{Lemma}
\newtheorem{coro}{Corollary}
\centerline{\bf BIAS-REDUCED EXTREME QUANTILE ESTIMATORS OF WEIBULL TAIL-DISTRIBUTIONS}
\vskip4ex
\large  Jean Diebolt$^{(1)}$, Laurent Gardes$^{(2)}$, St\'ephane Girard$^{(2)}$ and Armelle Guillou$^{(3)}$

\vskip4ex
\noindent
{\it $^{(1)}$ CNRS, Universit\'e de Marne-la-Vall\'ee\\
\'Equipe d'Analyse et de Math\'ematiques Appliqu\'ees\\
5, boulevard Descartes, Batiment Copernic\\
Champs-sur-Marne\\
77454 Marne-la-Vall\'ee Cedex 2, France}

\vskip2ex
\noindent
{\it $^{(2)}$ INRIA Rh\^one-Alpes, team Mistis,\\
Inovall\'ee, 655, av. de l'Europe, Montbonnot,\\
 38334 Saint-Ismier cedex, France}

\vskip2ex
\noindent
{\it $^{(3)}$ Universit\'e Paris VI\\
Laboratoire de Statistique Th\'eorique et Appliqu\'ee \\
Bo\^ \i te 158\\
175 rue du Chevaleret \\
75013 Paris, France}

\vskip4ex
\noindent
{\bf Abstract.} {\it In this paper, we consider the problem of estimating an 
extreme quantile  of a Weibull tail-distribution. The new extreme quantile 
estimator has a reduced bias compared to the more classical ones proposed in 
the literature. It is based on an exponential regression model that was 
introduced in Diebolt et al. (2006). The asymptotic normality of the extreme
quantile estimator is established.   
We also introduce an adaptive selection procedure to determine the number
of upper order statistics to be used.  
A simulation 
study as well as an application to a real data set are provided in order to prove the
efficiency of the above mentioned methods.}


\vskip8ex
\date{april 2006} 
\vfill\eject

\section{Introduction}

\noindent Let $X_1, ..., X_n$ be a sequence of independent and identically distributed (i.i.d.) random variables with distribution function $F$. 
In the present paper, we assume that $F$ is a Weibull tail-distribution, which means that
\begin{eqnarray}
1-F(x) = \exp(-H(x)) \qquad \mbox{with} \qquad H^{-1}(x) := \inf\{t: H(t) \geq x\}= x^{\theta} \ell(x),
\label{model}
\end{eqnarray}
\noindent
where $\theta >0$ denotes the Weibull tail-coefficient and $\ell$ is a slowly varying function at infinity satisfying
\begin{equation}
\label{convl}
{\ell(\lambda x) \over \ell(x)} \longrightarrow 1, \, \mbox{as}\, x \to \infty, \, \mbox{for all}\, \lambda > 0.
\end{equation}
\noindent
Based on the limited sample $X_1, ..., X_n$, the question is how to obtain a 
good estimate for a quantile of order $1-p_n$, $p_n\to0$ defined by
$$x_{p_n}=\inf\{y: F(y)\geq 1-p_n\},$$ 
such that the quantile to be estimated is situated on the 
border of or beyond the range of the data. Extrapolation outside the sample occurs for instance in 
reliability~(Ditlevsen, 1994), hydrology~(Smith, 1991),
and finance~(Embrechts et al., 1997).
\noindent
Beirlant et al. (1996) investigated this estimation problem and proposed the following estimator of~$x_{p_n}$:
\begin{equation}
\widetilde x_{p_n} = X_{n-k_n+1,n} \left({\log ({1 / p_n}) \over \log ({n / k_n})}\right)^{\widetilde \theta_n},
\label{GardesGirard}
\end{equation}
where $X_{1,n}\leq  ...\leq X_{n,n}$ denote the order statistics associated to 
the original sample and $\widetilde \theta_n$ is an estimator of $\theta$.  
One can use for instance 
the estimator introduced in Diebolt et al. (2006):
\begin{equation}
\label{thetaz}
\widetilde \theta_n = \frac{1}{k_n}\sum_{i=1}^{k_n} i\log(n/i)\left(\log (X_{n-i+1,n})-\log (X_{n-i,n})\right).
\end{equation}
We refer to Gardes and Girard (2005) for a study of the properties of~(\ref{GardesGirard}). 
In the preceding equations, $k_n$ denotes an intermediate sequence, i.e. a sequence such that $k_n \to \infty$ and ${k_n / n}\to 0$ as $n\to \infty$.
See Broniatowski (1993), Beirlant et al. (1995, 1996), Girard (2004), 
and Gardes and Girard (2006) for other contributions to the estimation
of~$\theta$ and Beirlant et al. (2006) for Local Asymptotic Normality (LAN) results.
Denoting $\tau_n=\log(1/p_n)/\log(n/k_n)$, the
estimator~(\ref{GardesGirard}) can be rewritten as 
$$
\widetilde x_{p_n} = X_{n-k_n+1,n}\, \tau_n^{\widetilde \theta_n}.
$$
It appears that the extreme quantile of order $1-p_n$ is estimated
through an ordinary quantile of order $1-k_n/n$ with a multiplicative
correction $\tau_n^{\widetilde \theta_n}$.

\noindent
It will appear in the next section that
$\widetilde x_{p_n}$ exhibits a bias depending on the rate of 
convergence to 1 of the ratio of the slowly varying function $\ell$ in 
(\ref{convl}). 
In order to quantify this bias,   
a second-order condition is required. This assumption can be expressed as follows:\\
\noindent
{\bf Assumption $(R_{\ell}(b, \rho))$.} {\it There exists a constant $\rho < 0$ and a rate function $b$ satisfying $b(x)\to 0$ as $x \to \infty$, such that for all $\varepsilon > 0$ and $1<A<\infty$, we have
$$\sup_{\lambda \in [1, A]} \left\vert {\log ({\ell(\lambda x)/ \ell(x)}) \over b(x) K_{\rho}(\lambda)}-1\right\vert \leq \varepsilon,\quad \mbox{for $x$ sufficiently large,}$$
\noindent
with ${K_{\rho}(\lambda) = \int_1^{\lambda} t^{\rho-1}dt.}$}

\noindent
It can be shown that necessarily $\vert b\vert$ is regularly varying with index $\rho$ (see e.g. Geluk and de Haan, 1987).
\noindent
In this paper, we focus on the case where the convergence~(\ref{convl}) is slow, and thus when the
bias term in $\widetilde \theta_n$ and therefore in
$\widetilde x_{p_n}$ is large. This situation is described by the following assumption:
\begin{equation}
\label{hypob}
x|b(x)|\to\infty \mbox{ as } x\to\infty,
\end{equation}
which is fulfilled by Gamma, Gaussian and ${\mathcal{D}}$ distributions,
see Table~\ref{tabex}. The ${\mathcal{D}}$ distribution is
an adaptation of Hall's class (Hall and Welsh, 1985) to
the framework of Weibull tail-distributions, see the appendix
for its definition.  
The methodology that we propose in order to reduce the bias of
$\widetilde x_{p_n}$ is to use the following regression model proposed by Diebolt et al. (2006) for the log-spacings of upper order statistics:
\begin{eqnarray}
Z_j &:=& j\, \log \left({n/ j}\right) \left(\log (X_{n-j+1,n})-\log (X_{n-j,n})\right) \nonumber\\
&=& \left(\theta + b\left(\log \left({n / k_n}\right)\right)\left({\log ({n/k_n})\over \log({n/j})}\right)\right)f_j + o_{\mathbb P}\left(b\left(\log \left({n / k_n}\right)\right)\right),
\label{modele}
\end{eqnarray}
\noindent
for $1\leq j \leq k_n$,
where $(f_1, ..., f_{k_n})$ is a vector of independent and standard exponentially distributed random variables and the $o_{\mathbb P}-$term is uniform in $j$.
\noindent
This exponential regression model is similar to the ones proposed by Beirlant 
et al. (1999, 2002) and Feuerverger and Hall (1999) in the case of 
Pareto-type distributions. 
\noindent
The model (\ref{modele}) allows us to generate bias-corrected estimates 
$\widehat \theta_n$ for $\theta$ through a Least-Square (LS) 
estimation of~$\theta$ and $b(\log ({n / k_n}))$.  
The resulting LS estimates are then the following:
\begin{eqnarray*}
\cases{\qquad \quad \, \, \, \widehat \theta_n = \overline{Z}_{k_n} - \widehat b\left(\log({n / k_n})\right) \overline x_{k_n}  & \cr
&\cr
\displaystyle\widehat b\left(\log({n / k_n})\right) = {\sum_{j=1}^{k_n} (x_j-\overline x_{k_n})Z_j \left/ \sum_{j=1}^{k_n} (x_j-\overline x_{k_n})^2\right.} }& \cr
\end{eqnarray*}
\noindent
where $x_j = {\log ({n/k_n})}/{\log ({n/j})}$, $\overline x_{k_n} = {1 \over k_n} \sum_{j=1}^{k_n} x_j$ and $\overline Z_{k_n} = {1 \over k_n} \sum_{j=1}^{k_n} Z_j$. 
\noindent
 The asymptotic normality of the LS-estimator $\widehat \theta_n$ is established in Diebolt et al. (2006). 
Now, in order to refine $\widetilde x_{p_n}$, we can use the additional information about the slowly varying function $\ell$ that is provided by the LS-estimates for $\theta$ and $b$. To this aim, condition $(R_{\ell}(b, \rho))$ is used to approximate the ratio $F^{-1}(1-p_n) / X_{n-k_n+1,n}$, noting that
$$X_{n-k_n+1,n} \stackrel{d}{=} F^{-1}(U_{n-k_n+1,n}),$$
with $U_{1,n} \leq ... \leq U_{n,n}$ the order statistics of a uniform $(0, 1)$ sample of size $n$,
\begin{eqnarray*}
{x_{p_n} \over X_{n-k_n+1,n}} &\stackrel{d}{=} & {F^{-1}(1-p_n) \over F^{-1}(U_{n-k_n+1,n})} \\
&= & {(-\log (p_n))^{\theta} \over (-\log(1-U_{n-k_n+1,n}))^{\theta}} \, {\ell(-\log (p_n)) \over \ell(-\log(1-U_{n-k_n+1,n}))}\\
&\stackrel{d}{=} & {(-\log (p_n))^{\theta} \over (-\log(U_{k_n,n}))^{\theta}} \, {\ell(-\log (p_n)) \over \ell(-\log(U_{k_n,n}))}\\
&{\simeq} & \left({\log ({1 / p_n}) \over \log ({n / k_n})}\right)^{\theta} \, \exp\left\{b\left(\log ({n / k_n})\right) {\left({\log ({1 / p_n}) \over \log ({n / k_n})}\right)^{\rho} - 1 \over \rho}\right\}.
\end{eqnarray*}
\noindent
The last step follows by replacing $U_{k_n,n}$ with $k_n/n$. Hence, we arrive at the following estimator for extreme quantiles
$$\widehat x_{p_n} = X_{n-k_n+1,n} \, \left({\log ({1 / p_n}) \over \log ({n/k_n})}\right)^{\widehat \theta_n} \, \exp\left\{\widehat b\left(\log ({n / k_n})\right) {\left({\log ({1 / p_n}) \over \log ({n / k_n})}\right)^{\widehat \rho_n} - 1 \over \widehat \rho_n}\right\},$$
or equivalently,
$$
\widehat x_{p_n} = X_{n-k_n+1,n} \, \tau_n^{\widehat\theta_n} 
\exp\left(\widehat b(\log(n/k_n)) K_{\widehat\rho_n}(\tau_n)\right).
$$
\noindent
Here, $\widehat\rho_n$ is an arbitrary estimator of $\rho$.  
It will appear in the next section (see Theorem~\ref{thun}(ii)) that, if $\tau_n$ 
converges to a constant value $\tau>1$, one can even choose
$\widehat\rho_n=\rho^\#$ a constant value, for instance 
the canonical value $\rho^\#=-1$, as suggested by Feuerverger and Hall (1999).
Note that the estimator~(\ref{GardesGirard}) can be seen as a
particular case of $\widehat x_{p_n}$ obtained 
by neglecting the bias-term.  
In the following, we use the LS-estimators of $\theta$ and $b$ defined 
previously.  
The study of the asymptotic properties of the extreme quantile estimators
$\widehat x_{p_n}$ and $\widetilde x_{p_n}$ is the aim of Section~\ref{asymp}.
An adaptive selection procedure for $k_n$ is also proposed.  
A simulation study as well as a real data set are provided in 
Sections~\ref{simul} and~\ref{realdata}. Proofs are
postponed to Section~\ref{proofs}.

\section{Bias-reduced extreme quantile estimator}
\label{asymp}

\noindent
The asymptotic normality of our bias-reduced extreme quantile estimator $\widehat x_{p_n}$ is established in the following theorem.
\begin{theo}
\label{thun}
Suppose (\ref{model}) holds together with ($R_{\ell}(b, \rho))$ and (\ref{hypob}). We assume that 
\begin{equation}
\label{C1}
k_n \to \infty,  {\sqrt {k_n} \over \log({n / k_n})} b\left(\log ({n / k_n})\right)\, \to \lambda \in \mathbb R, 
\end{equation}
and if $\lambda=0$, 
\begin{equation}
\label{C2}
{\sqrt {k_n} \over \log({n / k_n})} \to \infty \, \mbox{and}\, {\log^2 (k_n) \over \log({n / k_n})} \to 0.
\end{equation}
Under the additional condition that
\begin{eqnarray}
|\widehat \rho_n-\rho| \log (\tau_n) = O_{\mathbb P}(1),
\label{C3}
\end{eqnarray}
we have
\begin{description}
\item[(i)] if $\tau_n \to \infty$
$${\sqrt {k_n} \over \log({n / k_n})\log (\tau_n)}\left(\log (\widehat x_{p_n})- \log (x_{p_n})\right) \stackrel{d}{\longrightarrow} {\cal N}(0, \theta^2),$$
\item [(ii)] if $\tau_n \to \tau, \tau >1$, and if we replace $\widehat \rho_n$ by a canonical choice $\rho^{\#} <0,$ then
$${\sqrt {k_n} \over \log({n / k_n})}\left(\log (\widehat x_{p_n})- \log (x_{p_n})\right) \stackrel{d}{\longrightarrow} {\cal N}\left(\lambda\mu(\tau), \theta^2\sigma^2(\tau)\right),$$
with
$$\sigma^2(\tau)=\left(K_{\rho^\#}(\tau)-\log (\tau)\right)^2,$$
and
$$\mu(\tau) = \left(K_{\rho^\#}(\tau)-K_{\rho}(\tau)\right).$$
\end{description}
\end{theo}
\noindent
In the following remark we provide some possible choices for
the sequences $(k_n)$ and $(p_n)$.  
\begin{Remark} Suppose (\ref{model}) holds together with
($R_{\ell}(b, \rho)$) and (\ref{hypob}).
Then, choosing 
$$
k_n=\left(\lambda\frac{\log (n)}{b(\log (n))}\right)^2,\;\lambda\neq 0,\;
p_n= n^{-\tau},\; \tau>1, \mbox{ and } \widehat \rho_n=\rho^{\#} <0 ,
$$
Theorem~\ref{thun}(ii) applies and thus
$${1 \over b(\log (n))}\left(\log (\widehat x_{p_n})- \log (x_{p_n})\right) \stackrel{d}{\longrightarrow} {\cal N}\left(\mu(\tau), \left(\frac{\theta}{\lambda}\right)^2\sigma^2(\tau)\right).$$
Clearly, the faster $b$ converges to 0, the faster $\widehat x_{p_n}$
converges to $x_{p_n}$. 
\end{Remark}
\noindent
As a comparison, one can establish similar results for $\widetilde x_{p_n}$.  
\begin{theo}
\label{thdeux}
Suppose (\ref{model}) holds together with ($R_{\ell}(b, \rho))$ and (\ref{hypob}). We assume that 
$$
k_n \to \infty, \sqrt{k_n}b(\log(n/k_n)) \to \lambda \in {\mathbb R},
\lim\inf \tau_n>1,
$$
and if $\lambda=0$, $\log(k_n)/\log(n) \to 0$, we have:
\[ \frac{\sqrt{k_n}}{\log(\tau_n)} \left\{ \log (\widetilde x_{p_n}) - \log(x_{p_n})- b(\log(n/k_n))\left( \frac{\log(\tau_n)}{k_n} \sum_{j=1}^{k_n} \left ( \frac{\log (n/j)}{\log (n/k_n)} \right )^{\rho} - K_\rho(\tau_n) \right)\right\} 
\stackrel{d}{\longrightarrow} {\cal{N}}(0,\theta^2). \]
\end{theo}
The next corollary allows an asymptotic comparison of $\widehat x_{p_n}$
and $\widetilde x_{p_n}$.
\begin{coro}
Under the assumptions of Theorem~\ref{thdeux}, we have
\begin{description}
\item[(i)] if $\tau_n \to \infty$
$${\sqrt {k_n} \over \log (\tau_n)}\left(\log (\widetilde x_{p_n})- \log (x_{p_n})\right) \stackrel{d}{\longrightarrow} {\cal N}(\lambda, \theta^2),$$
\item [(ii)] if $\tau_n \to \tau, \tau >1$, then
$$\sqrt {k_n} \left(\log (\widetilde x_{p_n})- \log (x_{p_n})\right) \stackrel{d}{\longrightarrow} {\cal N}\left(\lambda\tilde\mu(\tau), \theta^2\tilde\sigma^2(\tau)\right),$$
with
$$\tilde\sigma^2(\tau)=\log^2 (\tau),$$
and
$$\tilde\mu(\tau) = \left(\log(\tau)-K_{\rho}(\tau)\right).$$
\end{description}
\end{coro}
In the situation where $\tau_n\to\infty$, clearly $\log(\widehat x_{p_n})$ is asymptotically
unbiased whereas $\log(\widetilde x_{p_n})$ is biased.
When $\tau_n\to\tau$, $\tau>1$ both estimates are asymptotically biased
but the bias of $\log(\widehat x_{p_n})$ can be smaller than the one
of $\log(\widetilde x_{p_n})$ if $\rho^\#$ is close to $\rho$.

\noindent Let us now introduce the empirical adapted 
 Asymptotic Mean Squared Error (AMSE$^*$) of $\widetilde x_{p_n}$ defined as
$$AMSE^*(\widetilde x_{p_n})  = 
\mbox{Asymptotic}\, \mathbb E\left(\log(\widetilde x_{p_n}) -\log(x_{p_n})\right)^2.$$
Note that we use an adapted version of the AMSE since it takes into account the fact that the distribution of the quantile estimators
is found to be closer to a lognormal distribution than to the asymptotic normal distribution. This is classical in the literature, see for instance  Matthys {\it et al.} (2004).
As a consequence of Theorem~\ref{thdeux}, we have
\begin{equation}
\label{AMSE}
AMSE^*(\widetilde x_{p_n}) = \theta^2 \frac{\log^2(\tau_n)}{k_n} + 
 b^2(\log(n/k_n)) 
\left\{ \frac{\log(\tau_n)}{k_n} \sum_{j=1}^{k_n} \left ( \frac{\log(n/j)}{\log(n/k_n)} \right )^{\rho} - 
K_{\rho}(\tau_n) \right \}^2. 
\end{equation}
We can now take benefit of the estimation of $b(\log(n/k_n))$ by
estimating the AMSE$^*$ given in~(\ref{AMSE}) by:
$$
\widehat{AMSE^*}(\widetilde x_{p_n})=
(\widehat \theta_n)^2 \frac{\log^2(\tau_n)}{k_n} + 
(\widehat b(\log(n/k_n)))^2 
\left\{ \frac{\log(\tau_n)}{k_n} \sum_{j=1}^{k_n} \left ( \frac{\log(n/j)}{\log(n/k_n)} \right )^{-1} + \tau_n^{-1}-1 \right\}^2. 
$$
Note that, in the latter formula, we replaced $\rho$ by a canonical choice ($\rho=-1$)
instead of estimating this parameter. In fact, this second-order parameter is
difficult to estimate in practice, and we can easily check by simulations that
fixing its value does not influence much the result (see e.g. Beirlant et al., 1999, 2002
or Feuerverger and Hall, 1999).
Then, the intermediate sequence $k_n$ can be selected by minimizing
the previous quantity: 
$$
\hat k_n=\arg\min_{k_n} \widehat{AMSE^*}(\widetilde x_{p_n}).
$$
This adaptive procedure for selecting the number of upper order
statistics is in the same spirit as the one proposed by
Matthys {\it et al.} (2004) in the case of the Weissman estimator (Weissman, 1978).  
In order to illustrate the usefulness of the bias reduction and
of the selection procedure, we provide a simulation study in the next section.

\section{A small simulation study}
\label{simul}

\begin{figure}[p]
\begin{tabular}{c c}
\subfigure[Median as a function of $k_n$]{\epsfig{figure=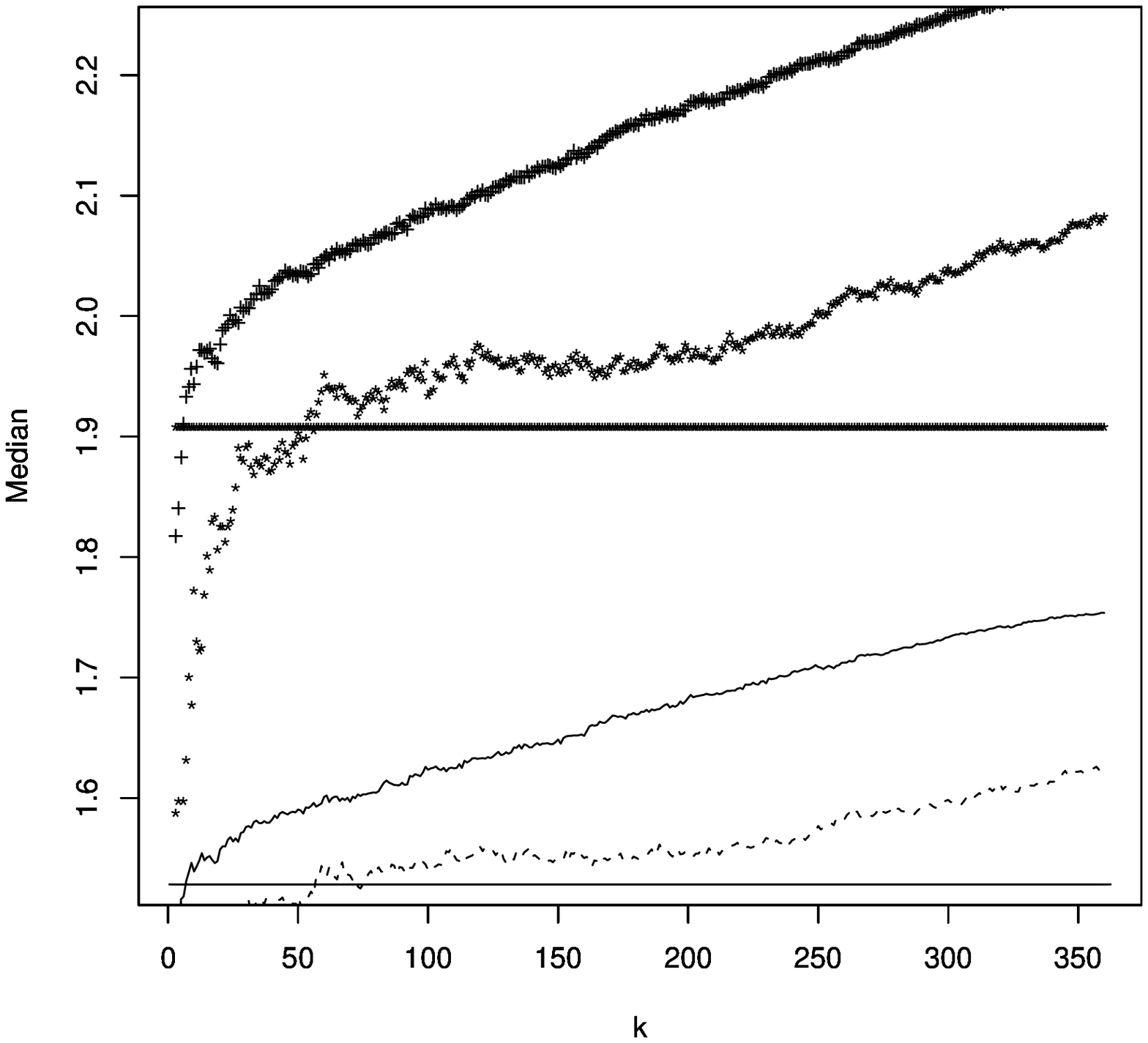,width=0.5\textwidth}} & 
\subfigure[MSE as a function of $k_n$]{\epsfig{figure=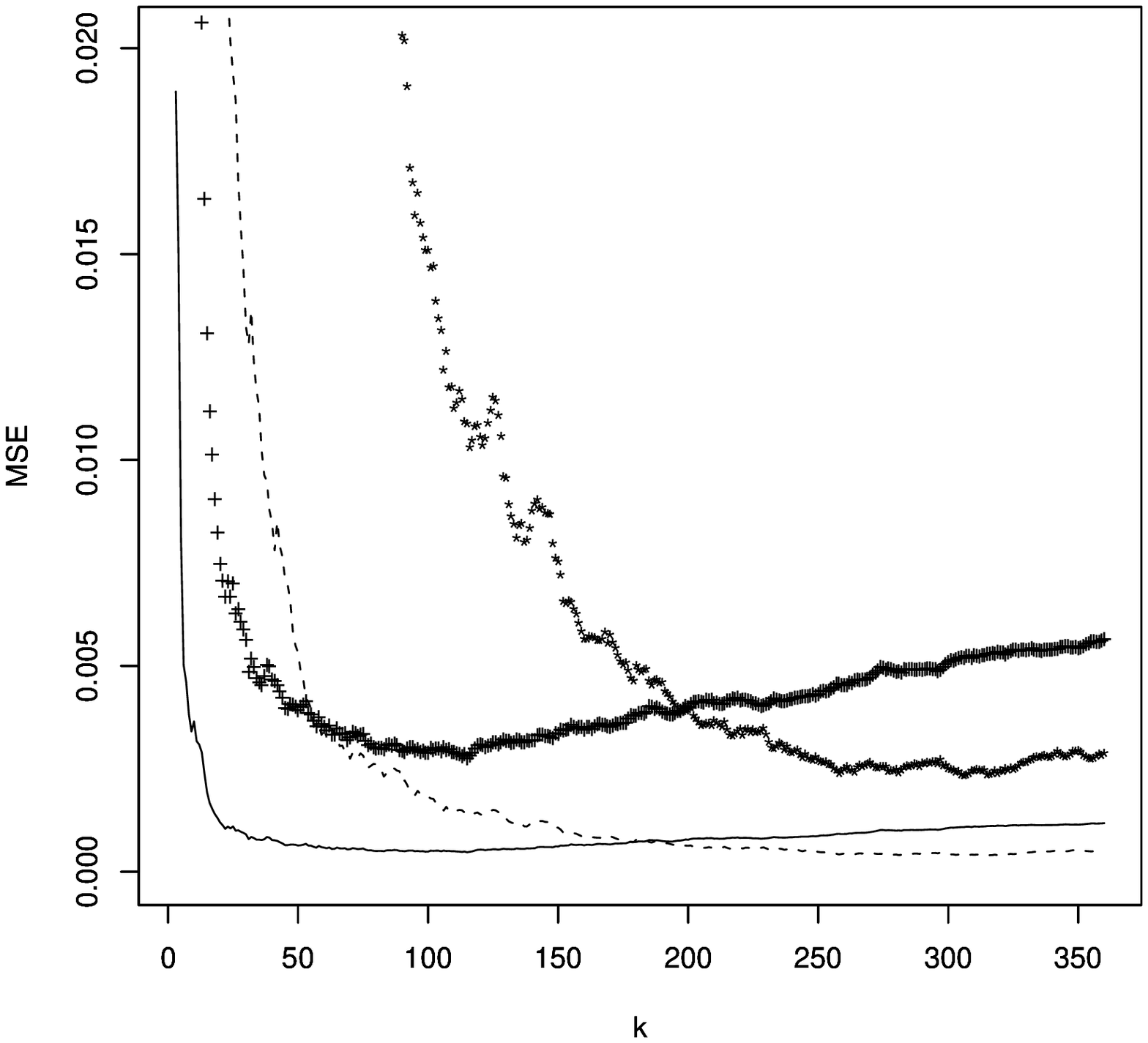,width=0.5\textwidth}} 
\end{tabular}
\caption{Comparison of 
$\log(\widetilde{x}_{p_n})$ ($\tau=2$: continuous line,
$\tau=4$: $+++$)
and $\log(\widehat{x}_{p_n})$ ($\tau=2$: dashed line,
$\tau=4$: $\star\star\star$)
for 
the $\left|{\cal{N}}(0,1)\right|$ distribution. The horizontal lines
on the left panel represent the true values of $\log(x_{p_n})$ 
(thin line: $\tau=2$, thick line: $\tau=4$).}
\label{fignorm}
\end{figure}

\begin{figure}[p]
\begin{tabular}{c c}
\subfigure[Median as a function of $k_n$]{\epsfig{figure=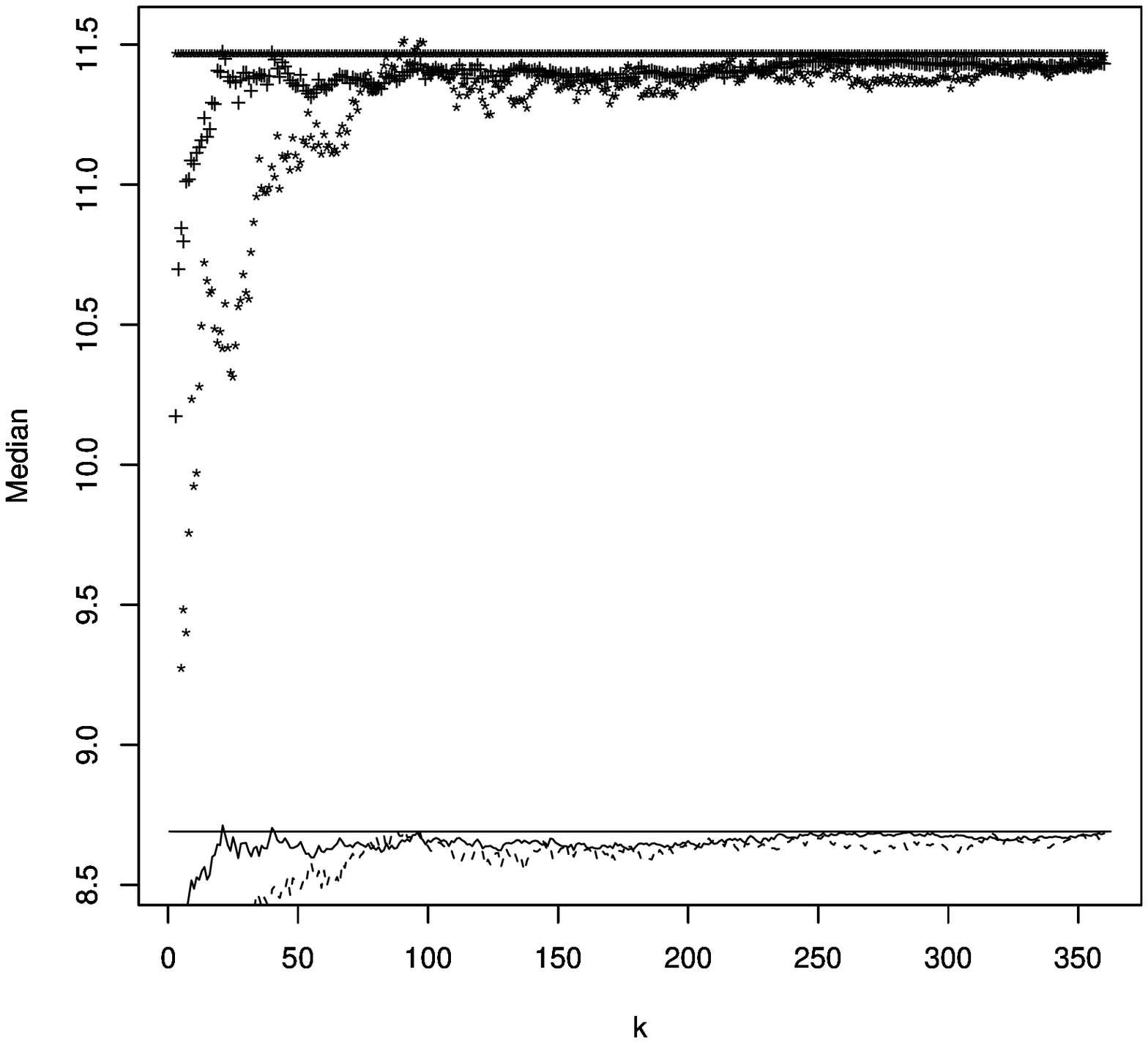,width=0.5\textwidth}} & 
\subfigure[MSE as a function of $k_n$]{\epsfig{figure=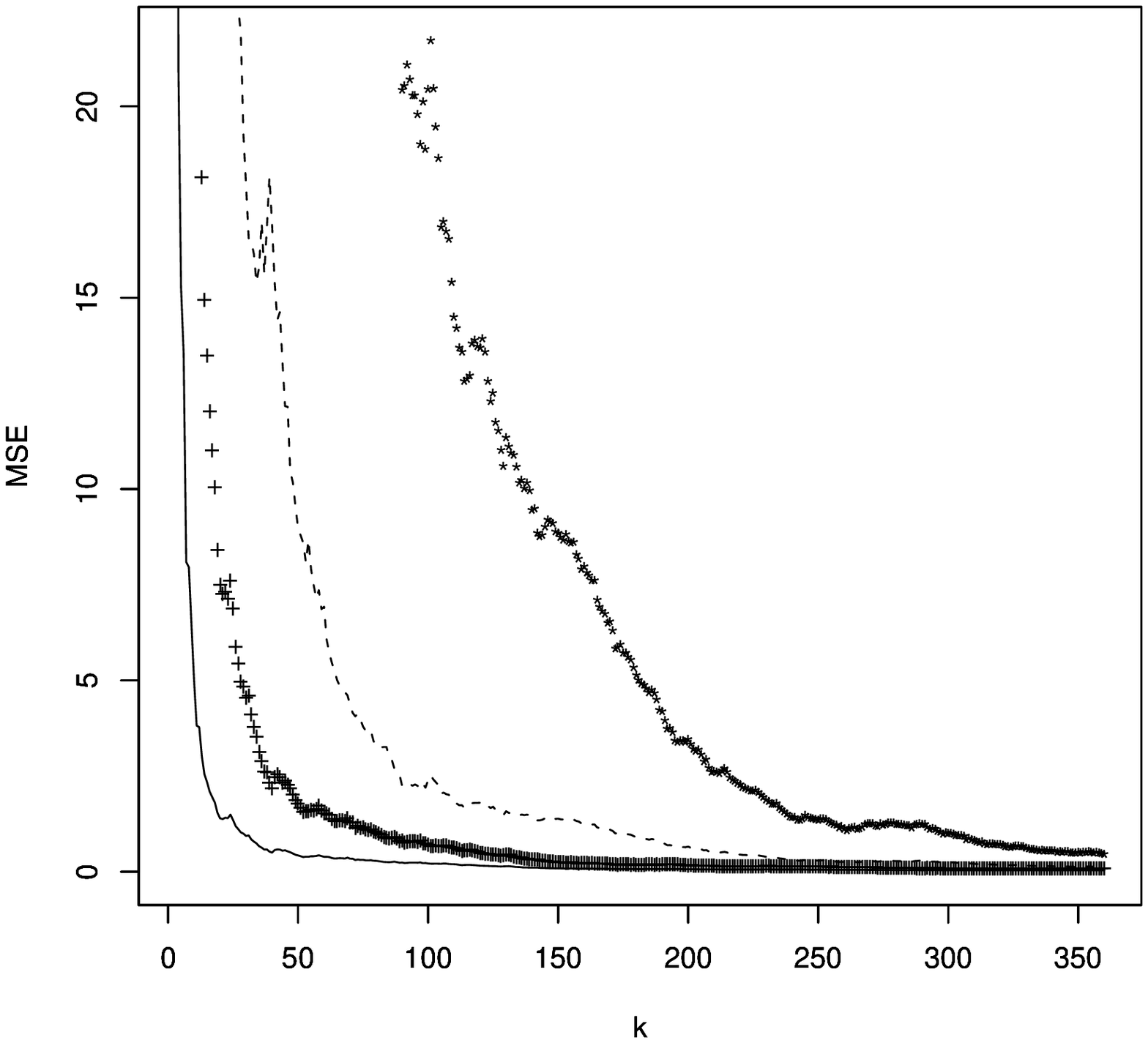,width=0.5\textwidth}} 
\end{tabular}
\caption{Comparison of 
$\log(\widetilde{x}_{p_n})$ ($\tau=2$: continuous line,
$\tau=4$: $+++$)
and $\log(\widehat{x}_{p_n})$ ($\tau=2$: dashed line,
$\tau=4$: $\star\star\star$)
for 
the ${\cal{W}}(0.25,0.25)$ distribution. The horizontal lines
on the left panel represent the true values of $\log(x_{p_n})$ 
(thin line: $\tau=2$, thick line: $\tau=4$).}
\label{figweib}
\end{figure}

\begin{figure}[p]
\begin{tabular}{c c}
\subfigure[Median as a function of $k_n$]{\epsfig{figure=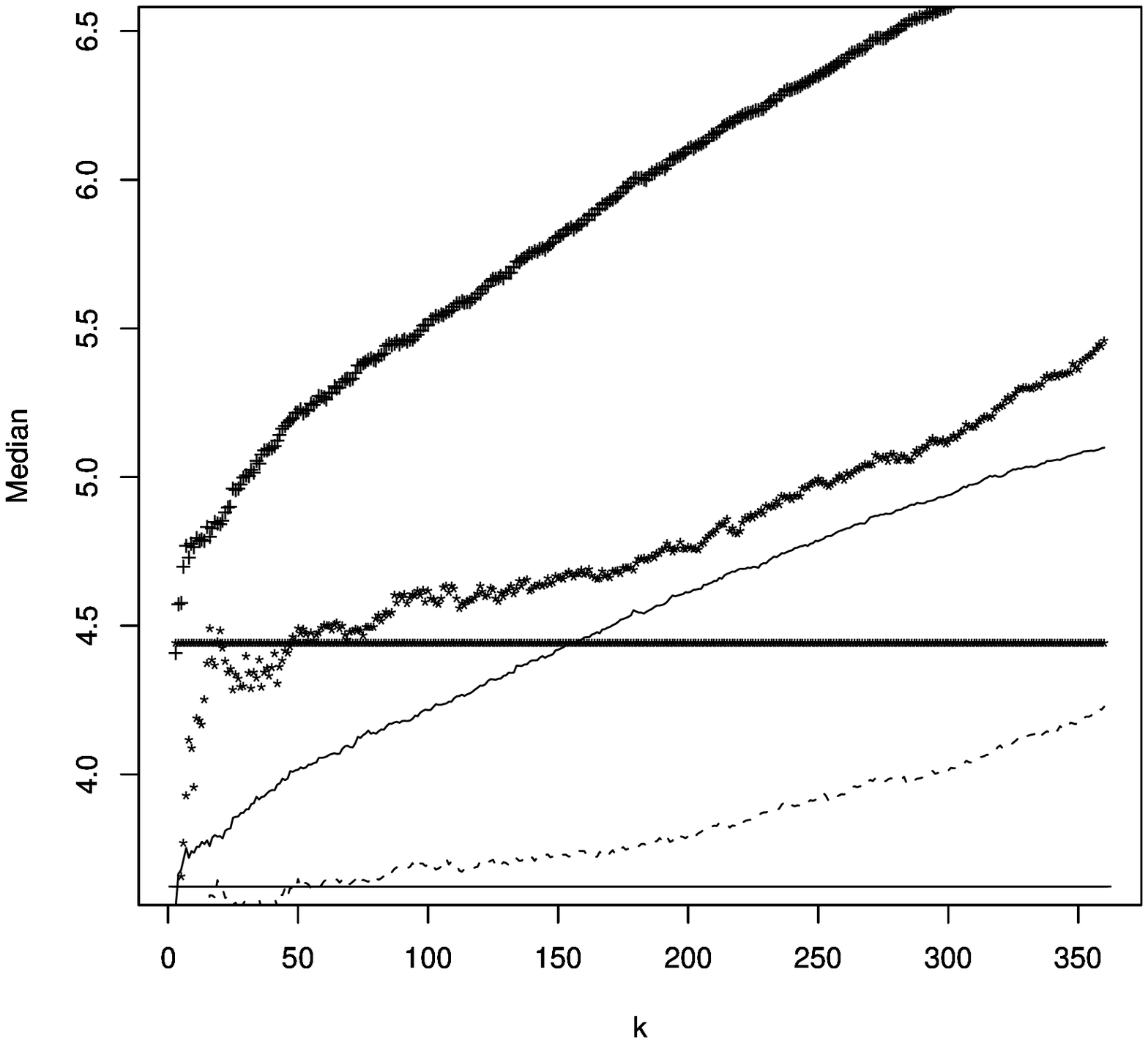,width=0.5\textwidth}} & 
\subfigure[MSE as a function of $k_n$]{\epsfig{figure=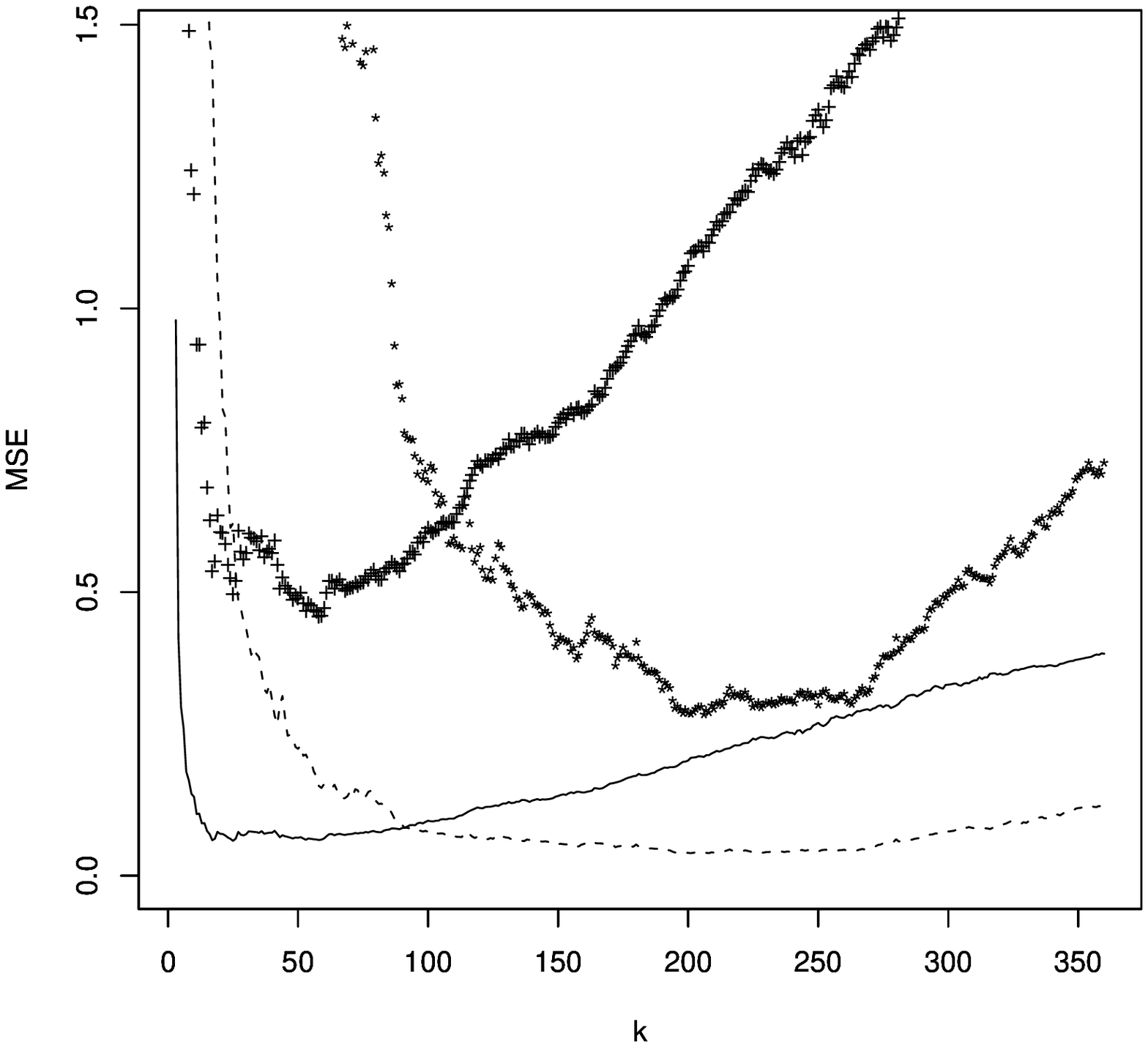,width=0.5\textwidth}} 
\end{tabular}
\caption{Comparison of 
$\log(\widetilde{x}_{p_n})$ ($\tau=2$: continuous line,
$\tau=4$: $+++$)
and $\log(\widehat{x}_{p_n})$ ($\tau=2$: dashed line,
$\tau=4$: $\star\star\star$)
for 
the $\Gamma(0.25,0.25)$ distribution. The horizontal lines
on the left panel represent the true values of $\log(x_{p_n})$ 
(thin line: $\tau=2$, thick line: $\tau=4$).}
\label{figgam1}
\end{figure}

\begin{figure}[p]
\begin{tabular}{c c}
\subfigure[Median as a function of $k_n$]{\epsfig{figure=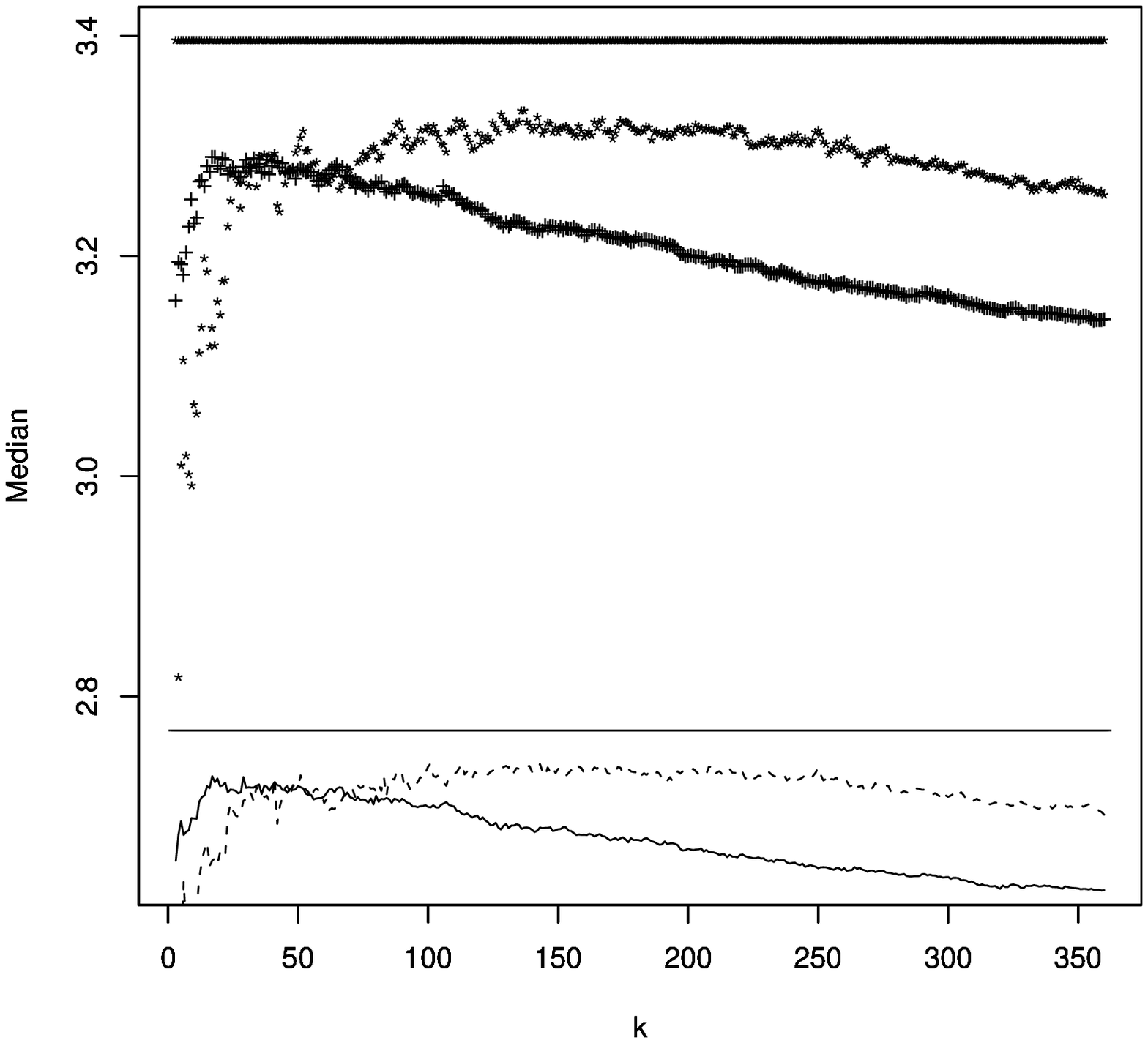,width=0.5\textwidth}} & 
\subfigure[MSE as a function of $k_n$]{\epsfig{figure=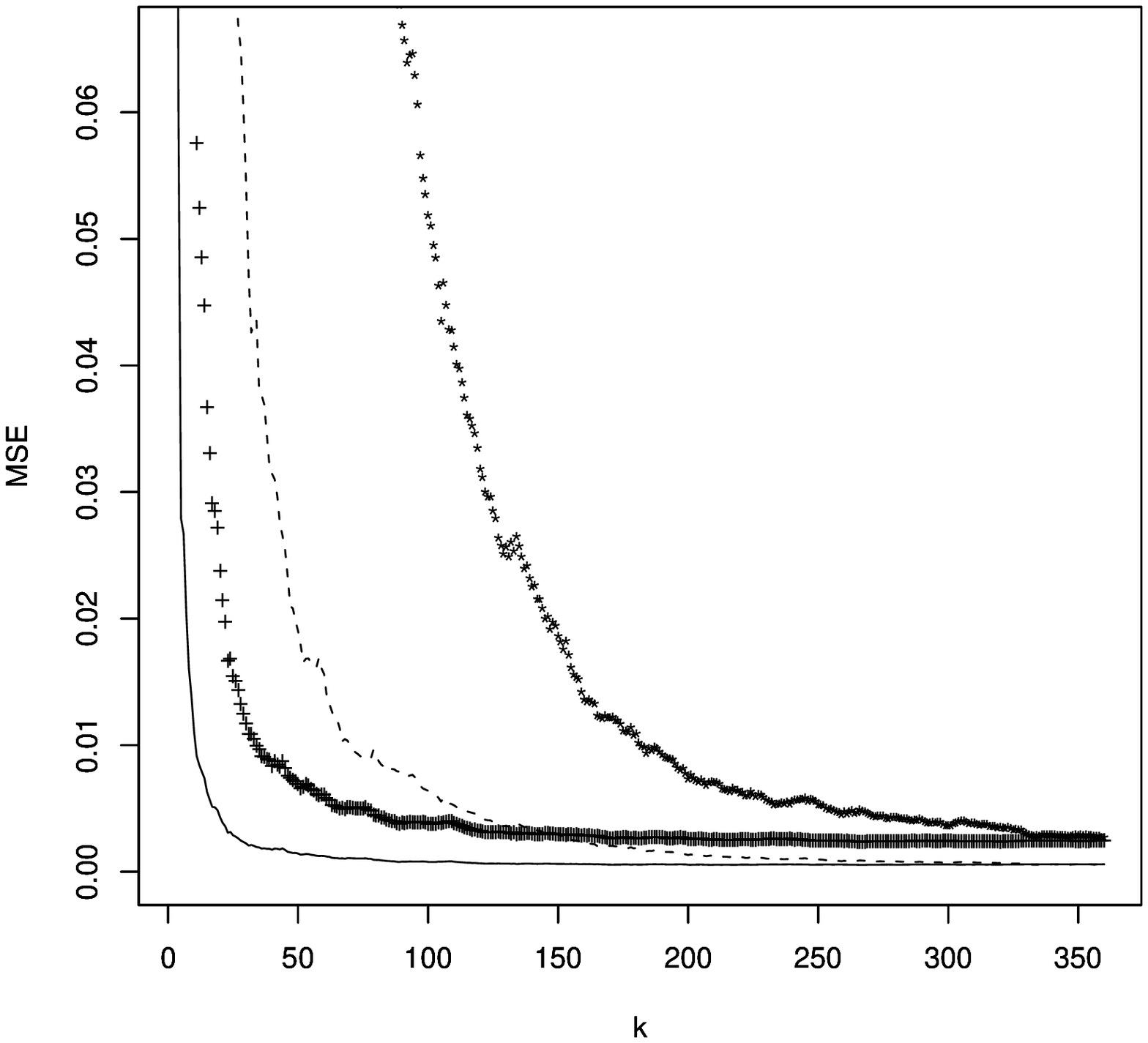,width=0.5\textwidth}} 
\end{tabular}
\caption{Comparison of 
$\log(\widetilde{x}_{p_n})$ ($\tau=2$: continuous line,
$\tau=4$: $+++$)
and $\log(\widehat{x}_{p_n})$ ($\tau=2$: dashed line,
$\tau=4$: $\star\star\star$)
for 
the ${\mathcal{D}}(1,0.5) $ distribution. The horizontal lines
on the left panel represent the true values of $\log(x_{p_n})$ 
(thin line: $\tau=2$, thick line: $\tau=4$).}
\label{figgam2}
\end{figure}

\noindent 
First, the finite sample performance of the estimators 
$\widetilde x_{p_n}$ and $\widehat x_{p_n}$ are investigated on 4 different distributions: $\left|{\cal{N}}(0,1)\right|$, $\Gamma(0.25,0.25)$, ${\mathcal{D}}(1,0.5)$ and ${\cal{W}}(0.25,0.25)$. It is shown in Gardes and Girard (2005) that $\widetilde x_{p_n}$ gives better results than the other approaches (Hosking and Wallis, 1987; Breiman et al., 1990; Beirlant et al., 1995). This explains why $\widehat x_{p_n}$ is only compared to the estimator $\widetilde x_{p_n}$.
In the following, we take $p_n := p_n(\tau)=n^{-\tau}$ with $\tau=2$ and 4 and 
we choose $\widehat \rho_n = -1$. We simulate $N=500$ samples 
$({\cal{X}}_{n,i})_{i=1,\ldots,N}$ of size $n=500$. On each sample 
$({\cal{X}}_{n,i})$, the estimates $\widehat x_{p_n(\tau),i}$, are computed
for $\tau=2$, $4$ and for $k_n=2,\ldots,360$. We present the plots obtained by 
drawing the points
\[ (k_n,\mbox{med}_i(\log(\widehat x_{p_n(\tau),i}))) \ {\rm{for}} \ \tau=2 \ {\rm{and}} \ 4, \]
where $\mbox{med}_i(\log(\widehat x_{p_n(\tau),i}))$ is the median value of 
$\log(\widehat x_{p_n(\tau),i})$, ${i=1,\ldots,N}$. 
We also present the associated MSE plots
$$
\left( k_n, \frac{1}{N}\sum_{i=1}^N \left(\log(\widehat x_{p_n(\tau),i})-\log(x_{p_n(\tau)})\right)^2\right)
\ {\rm{for}} \ \tau=2 \ {\rm{and}} \ 4. 
$$
The same procedure is achieved for the estimator $\widetilde x_{p_n}$.
Results are presented on figures~\ref{fignorm}--\ref{figgam2}. For the 
$\left|{\cal{N}}(0,1)\right|$, $\Gamma(0.25,0.25)$ and ${\mathcal{D}}(1,0.5)$ 
distributions, the bias of $\log(\widehat x_{p_n})$ is smaller than the one of 
$\log(\widetilde x_{p_n})$. Let us highlight that, for the latter distribution,
a significant bias reduction is obtained although $\widehat\rho_n\neq\rho$.  
Moreover, bias reduction is usually associated with an increase in 
variance. However, as illustrated in panel (b) of figures~\ref{fignorm}--\ref{figgam2},
our estimator $\widehat x_{p_n}$ is  still competitive in an adapted MSE 
sense.
Note that for the ${\cal{W}}(0.25,0.25)$ distribution, Theorem~\ref{thun} does not apply since $xb(x)=0$. In this case, 
the behavior of $\log(\widetilde x_{p_n})$ is slightly better.

\begin{figure}[p]
\begin{tabular}{c c}
\subfigure[$\tau=2$]{\epsfig{figure=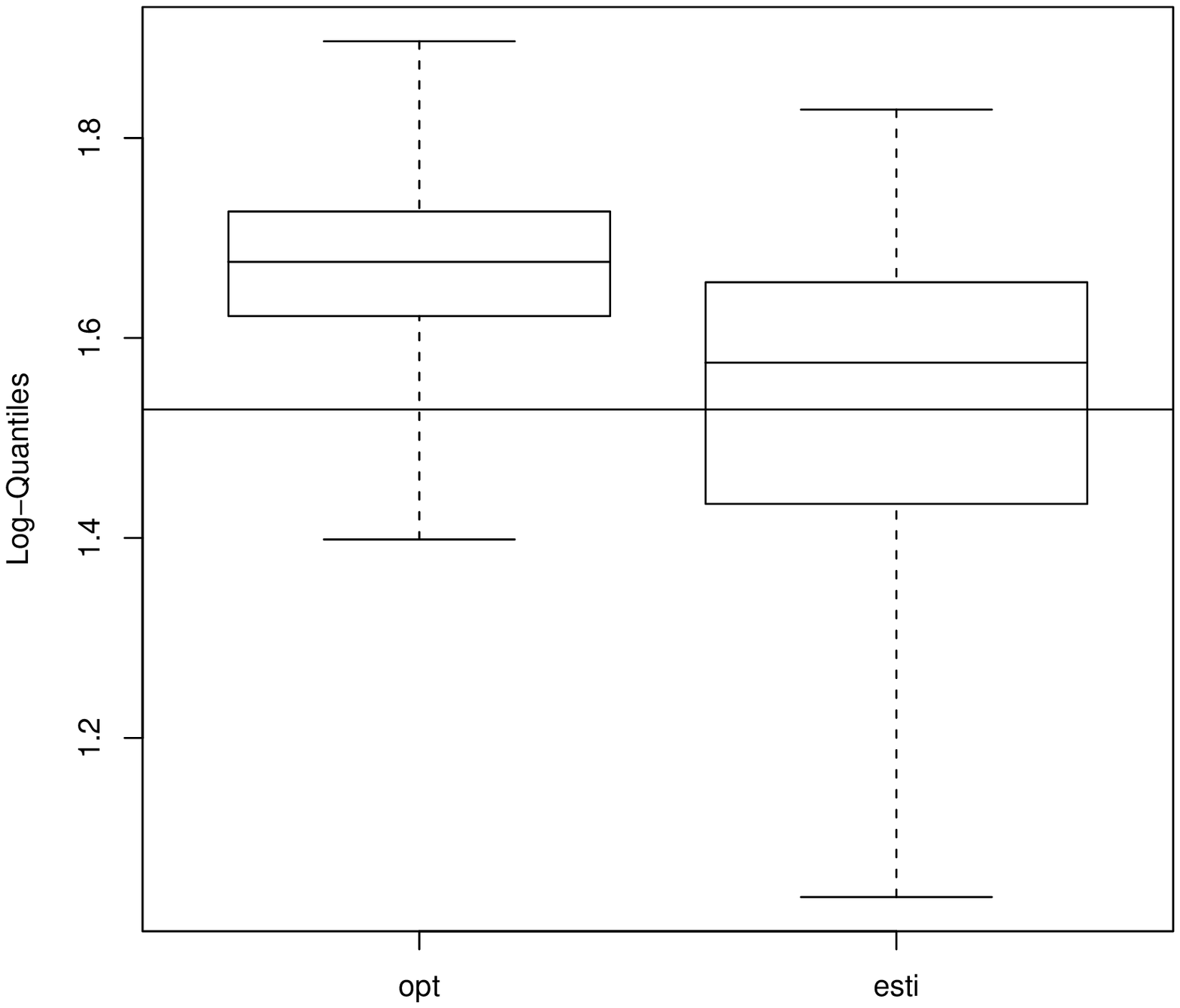,width=0.5\textwidth}} & 
\subfigure[$\tau=4$]{\epsfig{figure=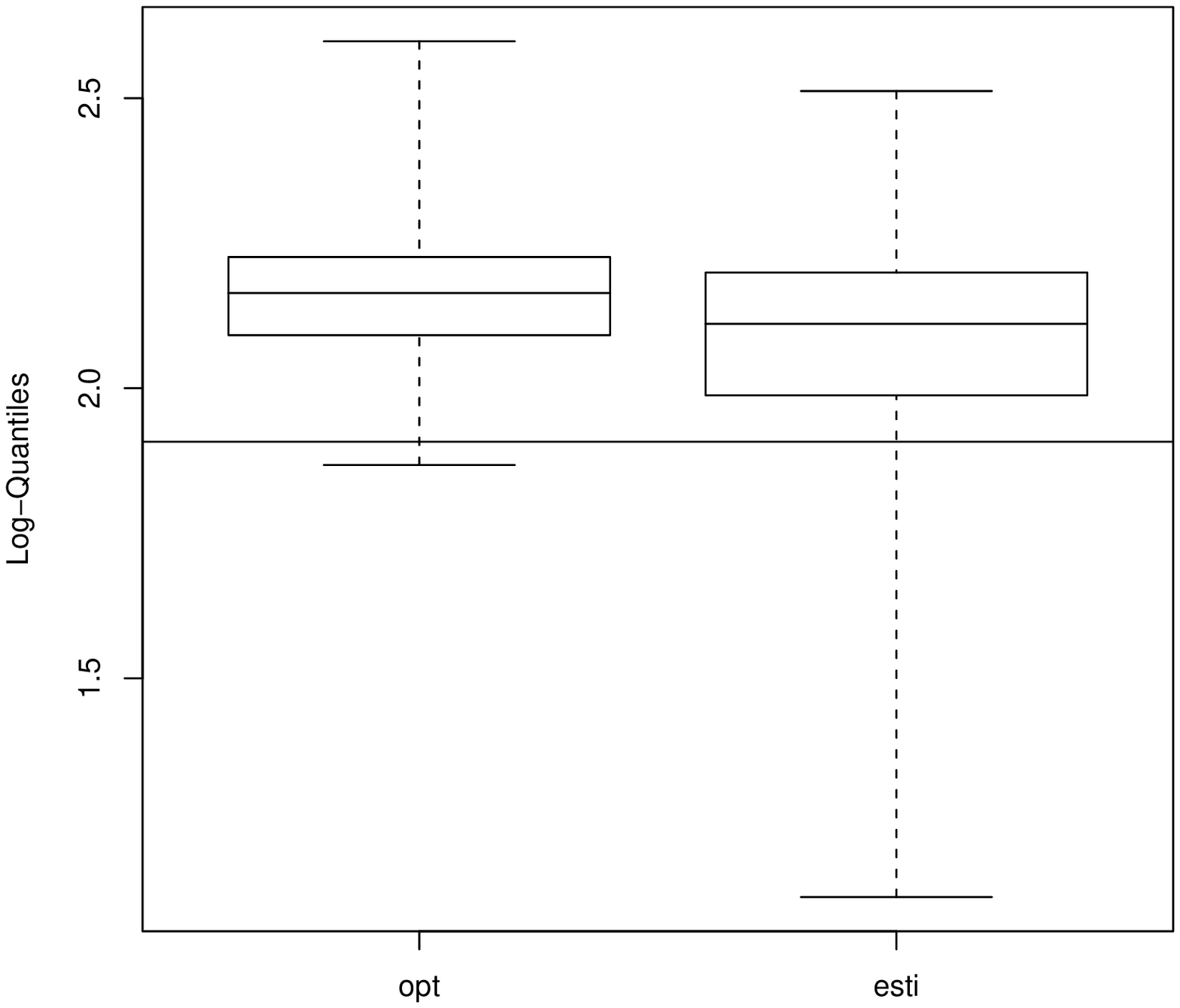,width=0.5\textwidth}} 
\end{tabular}
\caption{ $\left|{\cal{N}}(0,1)\right|$ distribution.
Boxplots of $\log(\widetilde x_{p_n})$ at the optimal value of $k_n$
obtained by minimizing the true AMSE$^*$ (left),
and at the value of $k_n$ obtained by minimizing the estimated AMSE$^*$
(right). The horizontal line indicates the true value
of $\log(x_{p_n})$.
}
\label{fignormbp}
\end{figure}

\begin{figure}[p]
\begin{tabular}{c c}
\subfigure[$\tau=2$]{\epsfig{figure=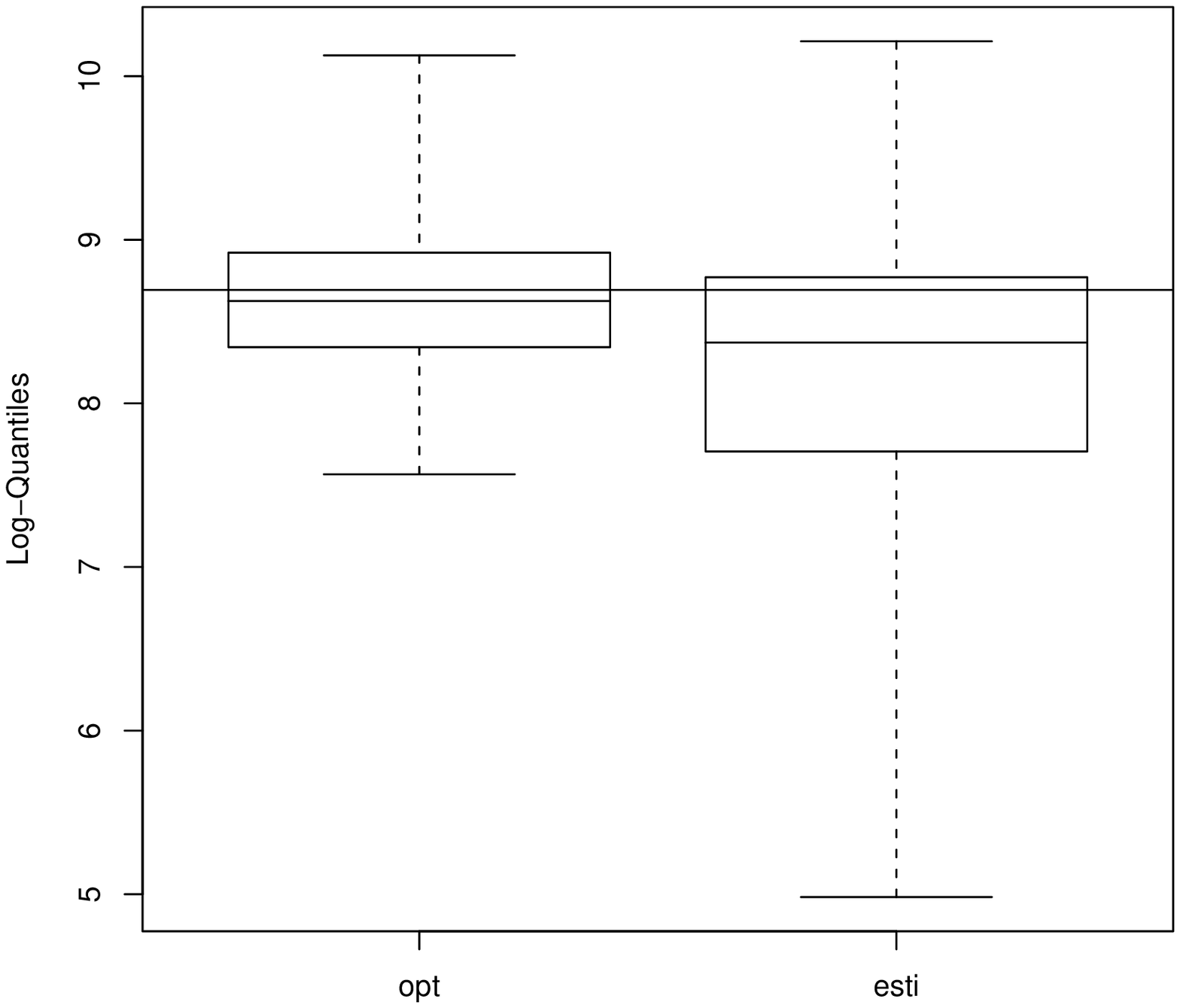,width=0.5\textwidth}} & 
\subfigure[$\tau=4$]{\epsfig{figure=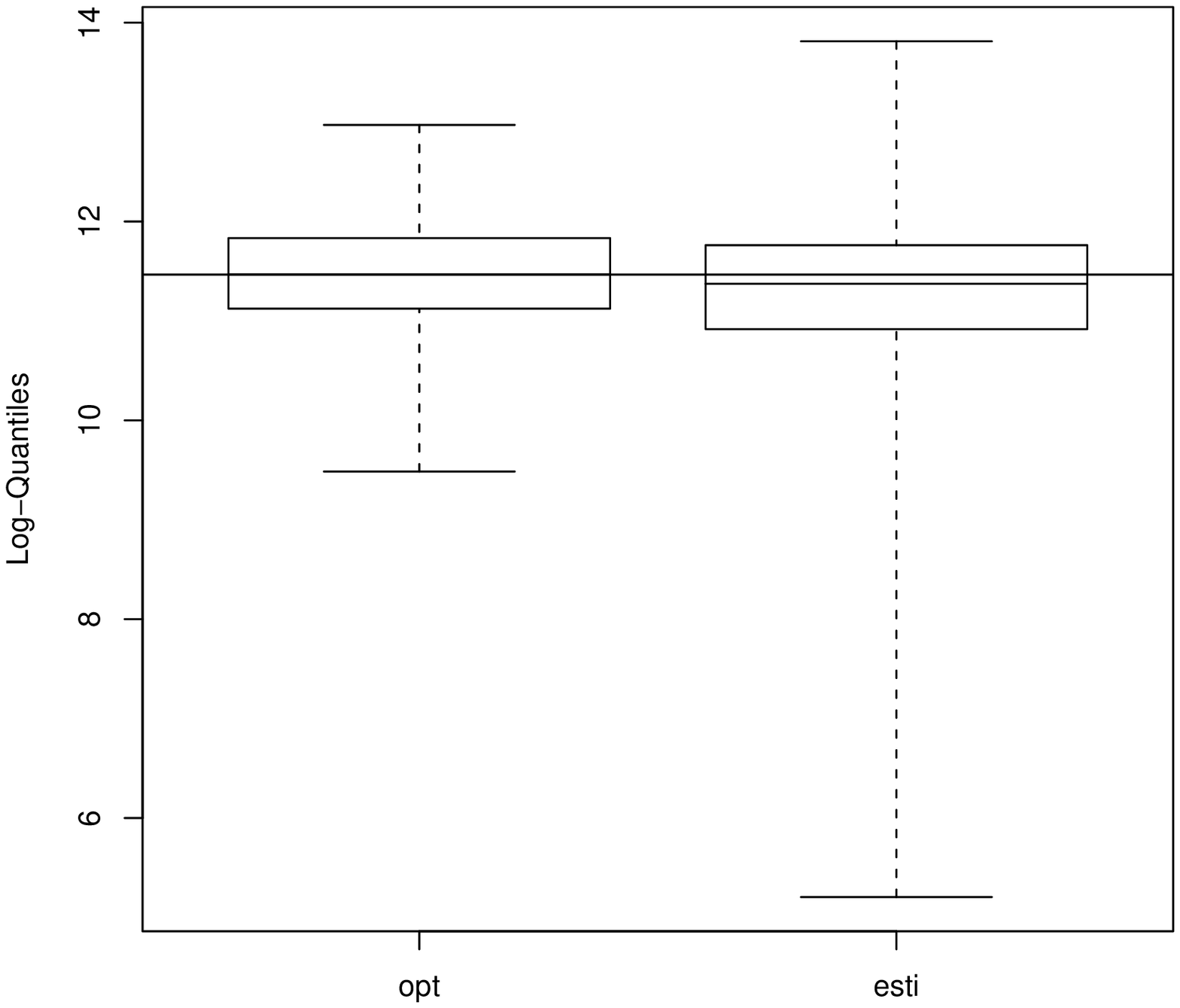,width=0.5\textwidth}} 
\end{tabular}
\caption{ ${\cal{W}}(0.25,0.25)$ distribution.
Boxplots of $\log(\widetilde x_{p_n})$ at the optimal value of $k_n$
obtained by minimizing the true AMSE$^*$ (left),
and at the value of $k_n$ obtained by minimizing the estimated AMSE$^*$
(right). The horizontal line indicates the true value
of $\log(x_{p_n})$.
}
\label{figweibp}
\end{figure}

\begin{figure}[p]
\begin{tabular}{c c}
\subfigure[$\tau=2$]{\epsfig{figure=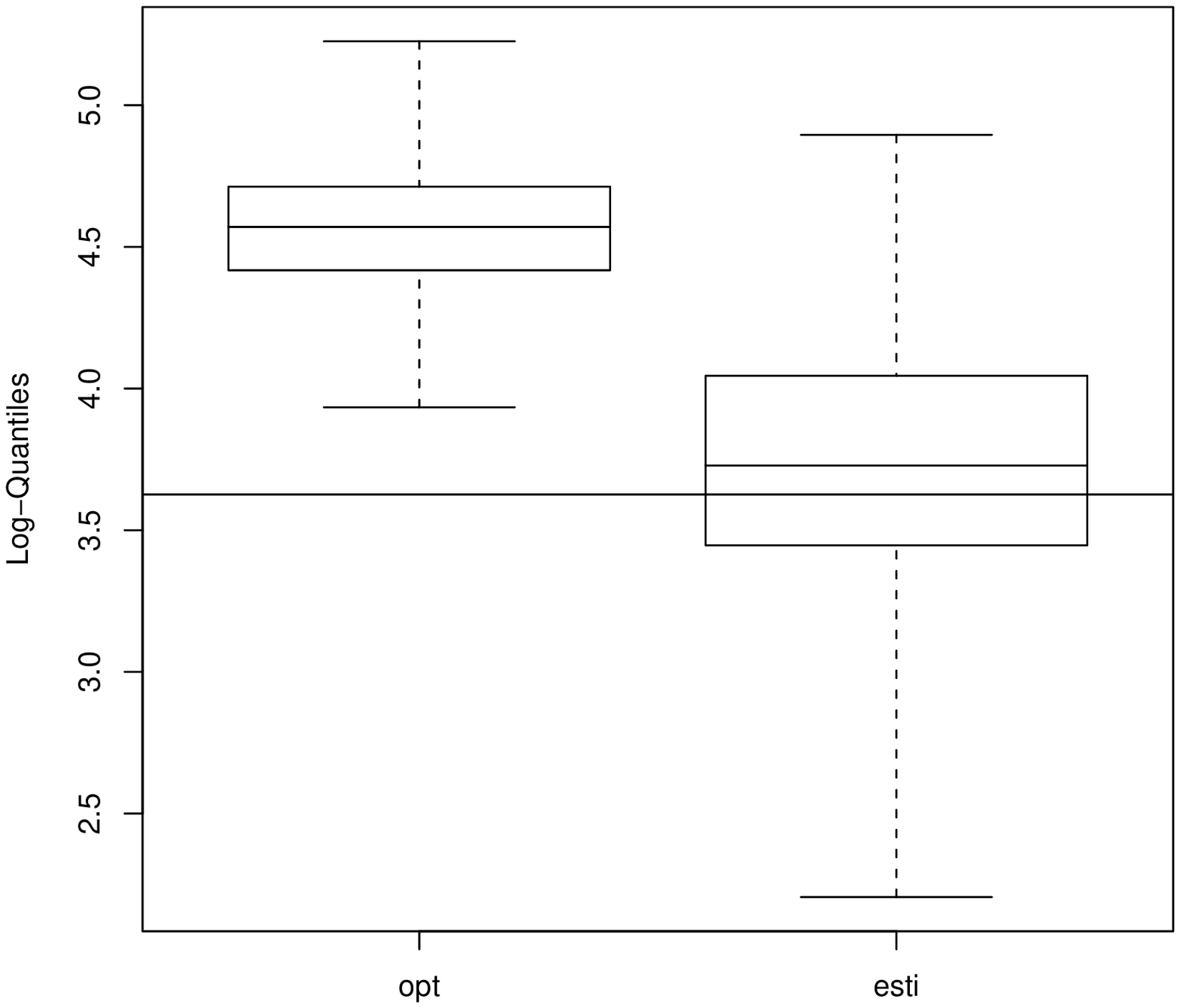,width=0.5\textwidth}} & 
\subfigure[$\tau=4$]{\epsfig{figure=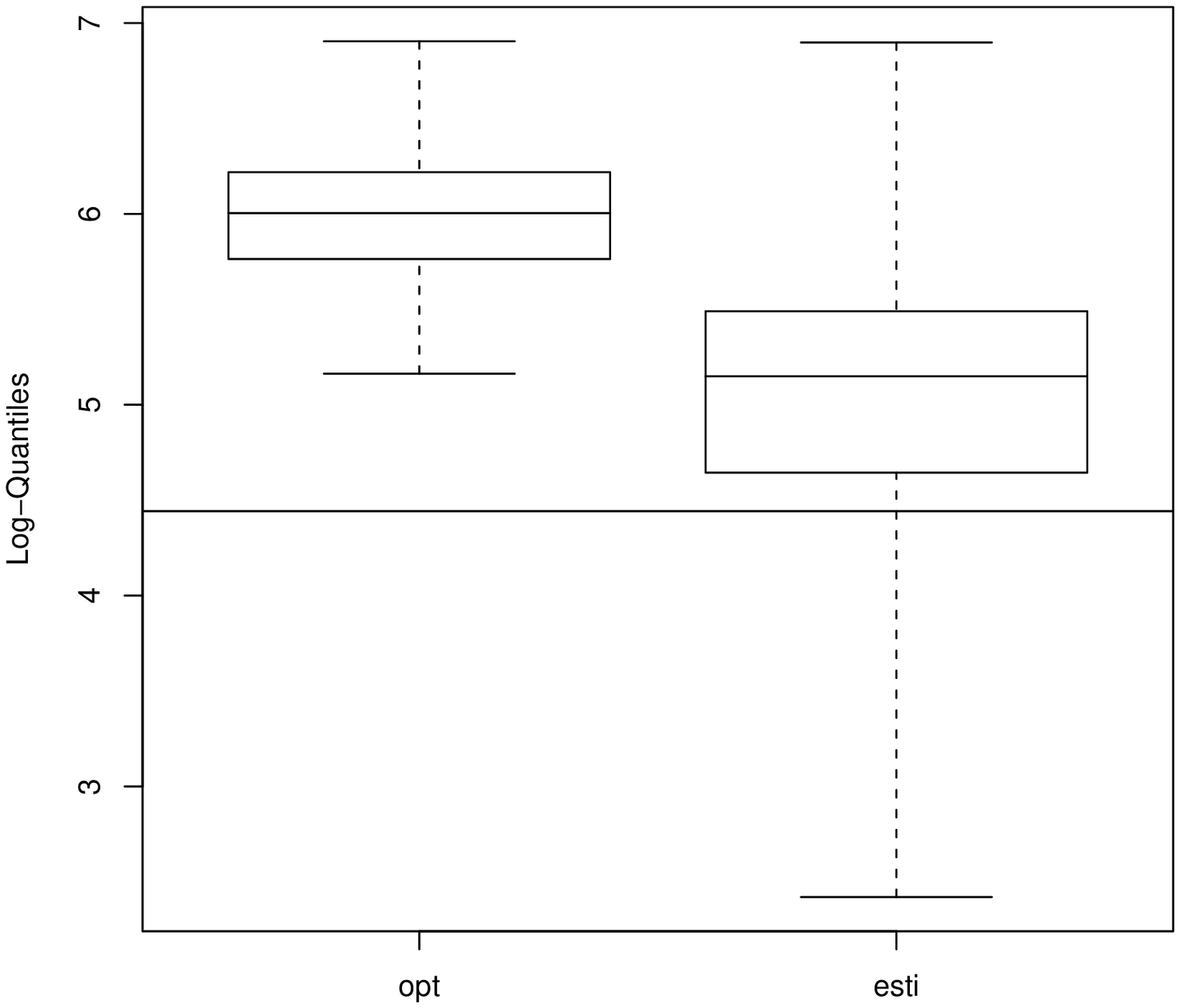,width=0.5\textwidth}} 
\end{tabular}
\caption{ $\Gamma(0.25,0.25)$ distribution.
Boxplots of $\log(\widetilde x_{p_n})$ at the optimal value of $k_n$
obtained by minimizing the true AMSE$^*$ (left),
and at the value of $k_n$ obtained by minimizing the estimated AMSE$^*$
(right). The horizontal line indicates the true value
of $\log(x_{p_n})$.
}
\label{figgam1bp}
\end{figure}

\begin{figure}[p]
\begin{tabular}{c c}
\subfigure[$\tau=2$]{\epsfig{figure=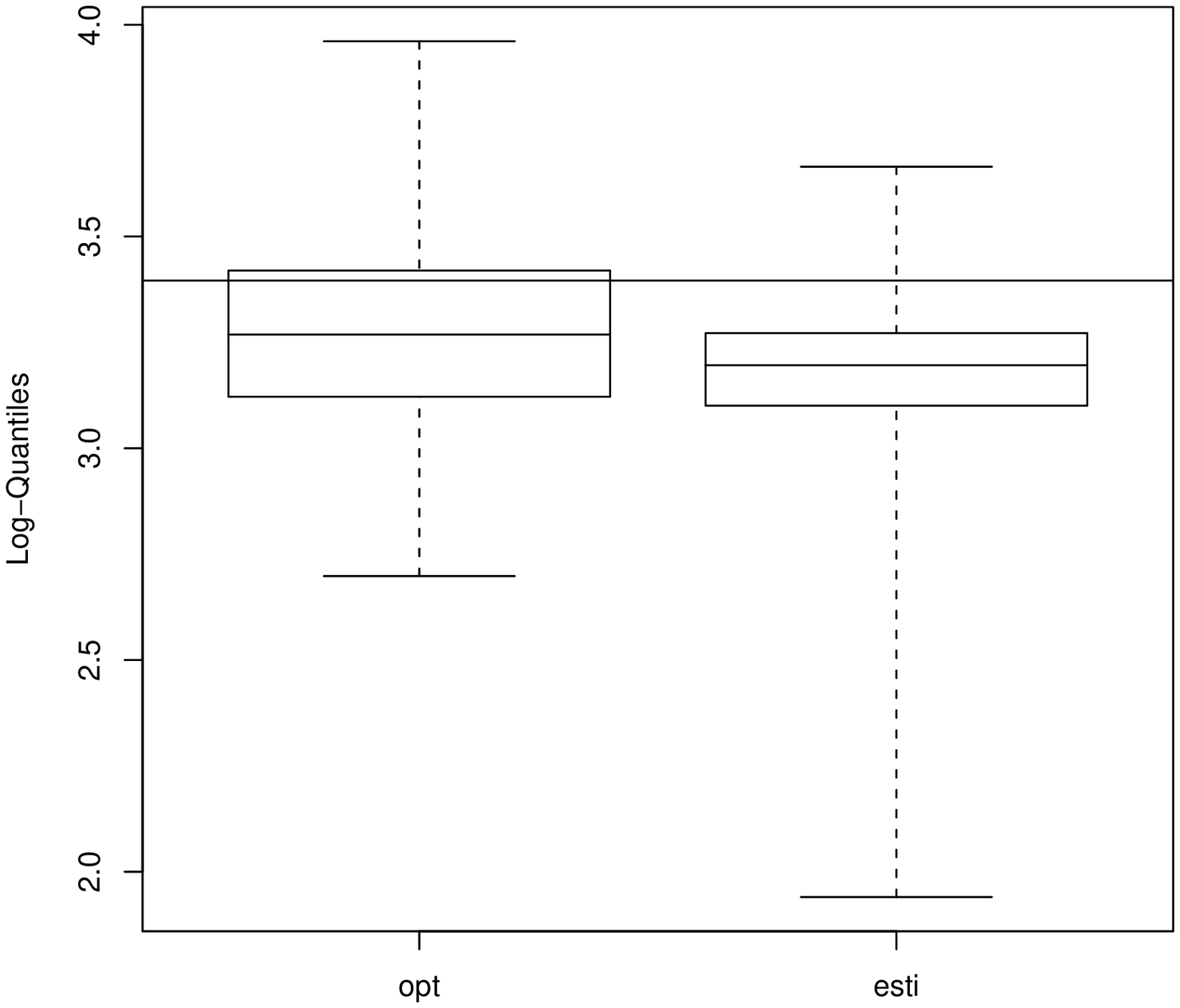,width=0.5\textwidth}} & 
\subfigure[$\tau=4$]{\epsfig{figure=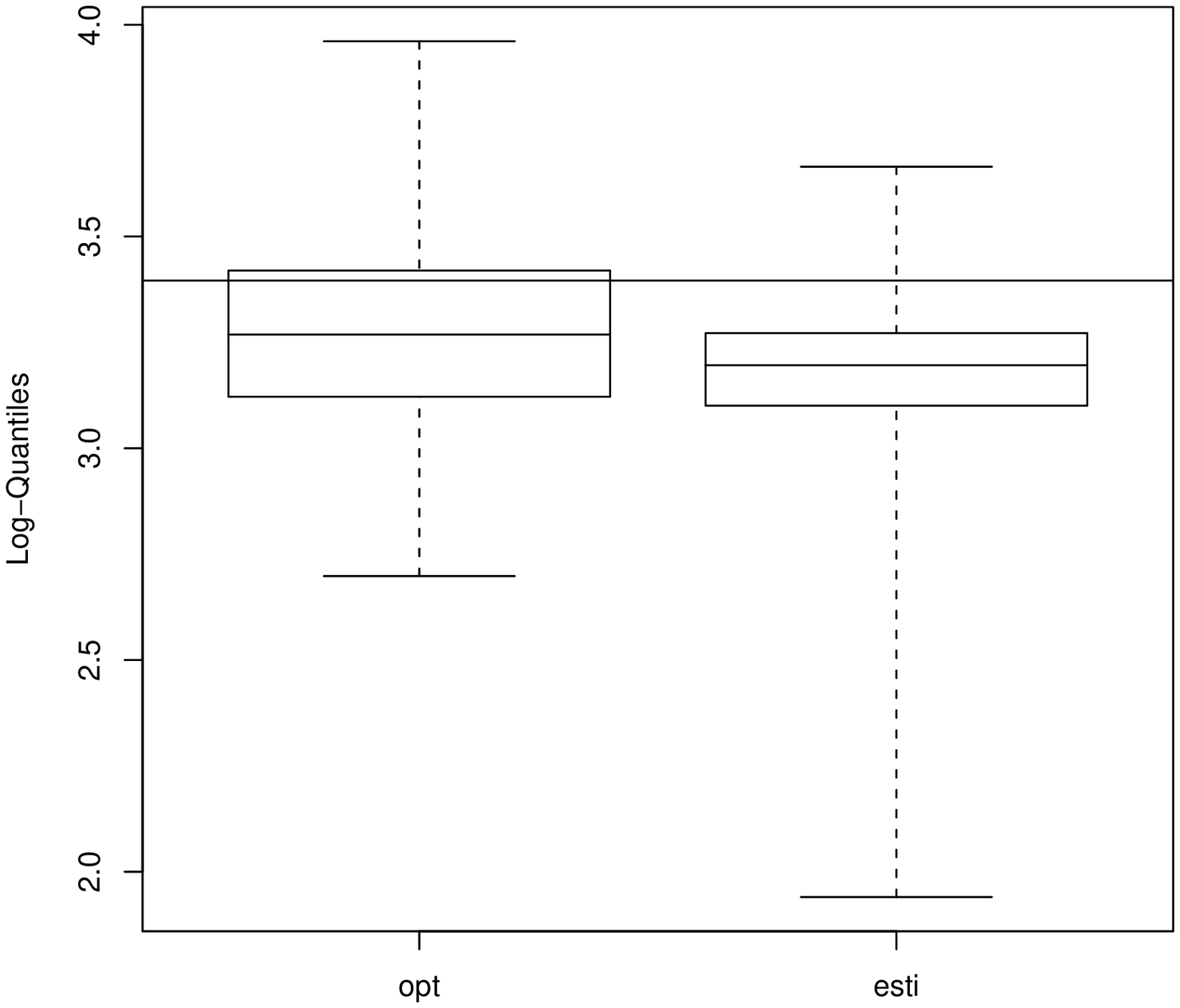,width=0.5\textwidth}} 
\end{tabular}
\caption{ ${\mathcal{D}}(1,0.5)$  distribution.
Boxplots of $\log(\widetilde x_{p_n})$ at the optimal value of $k_n$
obtained by minimizing the true AMSE$^*$ (left),
and at the value of $k_n$ obtained by minimizing the estimated AMSE$^*$
(right). The horizontal line indicates the true value
of $\log(x_{p_n})$.
}
\label{figgam2bp}
\end{figure}

\noindent Second, we investigate the behavior of the adaptive procedure
for selecting the number of upper order statistics in $\widetilde x_{p_n}$.
For $i=1,\dots,N$ and $\tau=2,4$, we denote by
$$
 {\widehat{k^{esti}_{n,i}}} = {\rm{arg}}\min_{k_n \in [1,n]}  \widehat{AMSE^*}(\widetilde x_{p_n(\tau),i}),
$$
the value selected on the sample $({\mathcal X}_{n,i})$.
As a comparison,
we introduce the value that would be obtained by
minimizing the true AMSE$^*$:
$$
{{k_{n}^{opt}}}=  {\rm{arg}}\min_{k_n \in [1,n]} {AMSE^*}(\widetilde x_{p_n(\tau)}).
$$
Figures~\ref{fignormbp}--\ref{figgam2bp}
contain the paired boxplots of the log-quantile estimators
$\log(\widetilde x_{p_n(\tau), i})$ at the adaptively selected values
${\widehat{k^{esti}_{n,i}}}$ on the right and at the sample fraction
${{k_n^{opt}}}$
with smallest AMSE$^*$ on the left. The horizontal line indicates the true
value of $\log(x_{p_n})$. The method proposed for the adaptive choice of
$k_n$,
while not achieving the goal of minimizing the MSE, does lead to
quantile estimators that exhibit good bias properties while the variance
is inflated in comparison to the asymptotically optimal values.

\section{Real data}
\label{realdata}

Here, the good performance of the adaptive selection procedure 
is illustrated through the analysis of extreme events
on the Nidd river data set.
This data set is standard in extreme value studies (see e.g. Hosking and Wallis, 1987,
or Davison and Smith, 1990.)
It consists in $154$ exceedances of the level~$65$
m$^3$s$^{-1}$ by the river Nidd (Yorkshire, England)
during the period 1934-1969 (35 years).
In environmental studies, the most common quantity of interest is
the $N$-year return level, defined as the level which is exceeded
on average once in~$N$ years.
Here, we focus on the estimation of the 50- and 100- year
return levels.
According to Hosking and Wallis (1987), the Nidd data may
reasonably be assumed to come from a distribution in the Gumbel
maximum domain of attraction.  This result was
confirmed in  Diebolt et al. (2006) who have shown that one could consider
Weibull tail-distributions as a possible model for such data.  
The Weibull tail-coefficient is estimated at $0.91$.   
We obtained $321.5 m^3s^{-1}$
as an estimation of the 50-year return level
and $359m^3s^{-1}$
as an estimation of the 100-year return level.
Note that these results are in accordance with the results
obtained by profile-likelihood or Bayesian methods, see Diebolt et al. (2005)
or Davison and Smith (1990).

\section{Proofs}
\label{proofs}

\noindent
{\bf Proof of Theorem~\ref{thun}.}  We decompose our quantile estimator
as follows:
\begin{eqnarray*}
\log (\widehat x_{p_n})-\log (x_{p_n}) &{=}& \log (X_{n-k_n+1,n}) + \widehat \theta_n \log (\tau_n)\\
&+& \widehat b\left(\log({n / k_n})\right) K_{\widehat\rho_n}(\tau_n) - \log\left((-\log (p_n))^{\theta}\ell(-\log (p_n))\right)\\ 
&\stackrel{d}{=}& \theta \left\{\log\left(-\log\left(U_{k_n,n}\right)\right)-\log\log({{n}/{k_n}})\right\}\\
&+& \left(\widehat \theta_n-\theta\right)\log (\tau_n) \\
&+&\left\{\log \ell\left(-\log\left(U_{k_n,n}\right)\right)-\log \ell\left(\log({n/ {k_n}})\right)\right\}\\
&+&\left\{\log \ell\left(\log({n/ {k_n}})\right) - \log \ell\left(-\log (p_n)
\right) + b\left(\log({n / k_n})\right)K_\rho(\tau_n)  \right\}\\
&+&\left(\widehat b\left(\log({n / k_n})\right)-b\left(\log({n / k_n})\right)\right) K_{\widehat\rho_n}(\tau_n) \\
&+&b\left(\log({n / k_n})\right) \left\{K_{\widehat \rho_n}(\tau_n) -K_{\rho}(\tau_n) \right\}\\
&:=& \sum_{j=1}^6 B_{j,k_n}.
\end{eqnarray*}
\noindent
We successively discuss each
of the terms $B_{j,k_n}$, $j=1, ..., 6.$ First concerning $B_{1,k_n}$, remark that
$$
\log\left(-\log (U_{k_n,n})\right) - \log\log({{n}/ {k_n}}) 
\stackrel{d}{=} \log \left ({T_{n-k_n+1,n} \over \log ({n / k_n})} \right ), 
$$
where $T_{j,n}$ denotes the order statistics from an i.i.d. standard exponential sample of size $n$. Since it is well known that
\begin{equation}
\label{convloiexp}
\sqrt {k_n} \left(T_{n-k_n+1,n}-\log ({n}/k_n)\right) \stackrel{d}{\longrightarrow} {\cal N}(0,1),
\end{equation}
we clearly have
\begin{eqnarray}
{\sqrt {k_n} \over {\log ({n / k_n}) \log (\tau_n)}} \, B_{1,k_n} = O_{\mathbb P}\left({1 \over \log^2 ({n / k_n}) \log \tau_n}\right) = o_{\mathbb P}(1).
\label{terme1}
\end{eqnarray}
Remark now that
\begin{eqnarray}
{\sqrt {k_n} \over {\log ({n / k_n}) \log (\tau_n)}} \, B_{2,k_n} = {\sqrt {k_n} \over {\log ({n / k_n})}} \left(\widehat \theta_n - \theta\right)\stackrel{d}{\longrightarrow} {\cal N}(0, \theta^2),
\label{terme2}
\end{eqnarray}
by Theorem~3.1 in Diebolt et al. (2006).
\noindent
Next, using $(R_{\ell}(b, \rho))$ and (\ref{convloiexp}), we get
\begin{eqnarray}
B_{3,k_n}
&\stackrel{d}{=}& \log 
\left(\frac{\ell(T_{n-k_n+1,n})}{\ell(\log(n/k_n))}\right)\nonumber\\
&{=}& K_\rho\left(\frac{T_{n-k_n+1,n}}{\log(n/k_n)}\right) b(\log (n/k_n))(1+o_{\mathbb P}(1))\nonumber\\
&{=}&\left(\frac{T_{n-k_n+1,n}}{\log(n/k_n)}-1\right) b(\log (n/k_n))(1+o_{\mathbb P}(1))\nonumber\\
&=&O_{\mathbb P}\left(\frac{b(\log (n/k_n))}{\sqrt{k_n}\log (n/k_n)}\right),
\label{terme3-0}
\end{eqnarray}
so that under (\ref{C1}),
\begin{eqnarray}
{\sqrt {k_n} \over {\log ({n / k_n}) \log (\tau_n)}} \, B_{3,k_n} = O_{\mathbb 
P}\left({1 \over \sqrt{k_n}\log(n/k_n)\log (\tau_n)}\right) = o_{\mathbb P}(1).
\label{terme3}
\end{eqnarray}
\noindent 
Next, under $(R_{\ell}(b, \rho))$ one has (for a suitably
chosen $b$) that for all $\epsilon >0$ (see Drees, 1998, Lemma
2.1)
$$ \sup_{t>1} t^{-(\epsilon + \rho)} \left\vert
{\log \ell(t x)-\log \ell(x) \over b(x)} -
K_{\rho}(t)\right\vert \longrightarrow 0.
$$
Hence, we conclude, choosing $\epsilon < -\rho$, that
\begin{equation}
\label{terme4-0}
\left\vert {B_{4,k_n} \over b(\log({n / k_n}))} \right\vert = \left\vert
 {\log \ell\left({\log (n /k_n)}\right)-\log \ell(\tau_n\log (n/k_n)) \over b(\log({n / k_n}))} + K_{\rho}(\tau_n)\right\vert \longrightarrow 0,
\end{equation}
\noindent
which implies that, under  (\ref{C1}), we have
\begin{eqnarray}
{\sqrt {k_n} \over {\log ({n / k_n}) \log (\tau_n)}} \, B_{4,k_n} = o\left({1 \over \log (\tau_n)}\right)=o(1).
\label{terme4}
\end{eqnarray}
\noindent
Next, we can check that, according to (\ref{modele}),
$$B_{5,k_n} = K_{\widehat \rho_n}(\tau_n) \, {1 \over k_n} \sum_{j=1}^{k_n} \beta_{j,n} (f_j-1),$$
\noindent
where 
$$\beta_{j,n} := {(x_j-\overline x_{k_n})(\theta+b(\log({n / k_n}))x_j) \over {1\over k_n} \sum_{i=1}^{k_n}(x_i-\overline x_{k_n})^2}.$$
A direct application of Lyapounov's theorem, combined with Lemma~7.3 in Diebolt et al. (2006) yields
$${\sqrt {k_n} \over \log({n / k_n})} {1\over k_n}\sum_{j=1}^{k_n} \beta_{j,n} (f_j-1) \stackrel{d}{\longrightarrow} {\cal N}(0, \theta^2).$$
\noindent
Therefore
\begin{eqnarray}
{\sqrt {k_n} \over {\log ({n / k_n}) \log (\tau_n)}} \, B_{5,k_n} = {K_{\widehat \rho_n}(\tau_n) \over \log (\tau_n)} \xi_{1,n} \quad \mbox{with} \quad \xi_{1,n} \stackrel{d}{\longrightarrow} {\cal N}(0, \theta^2).
\label{terme5}
\end{eqnarray}
\noindent
Finally, following the method of proof of Lemma~1 in de Haan and Rootz\'en
(1993), we find that 
$$\left\vert \int_1^{\tau_n} s^{\rho-1}\, (s^{\widehat \rho_n-\rho}-1)\, ds - (\widehat \rho_n-\rho)
\int_1^{\tau_n} s^{\rho-1} \, \log (s) \, ds \right\vert$$
\begin{eqnarray*}
&\leq &\vert \widehat \rho_n-\rho\vert \int_1^{\tau_n} s^{\rho-1} \, \log (s) \, (s^{\vert \widehat \rho_n -\rho \vert}-1)\, ds\\
&\leq& \vert \widehat \rho_n-\rho\vert \left(\int_1^{\tau_n} s^{\rho-1}\,  \log (s) \, ds\right)
\left(\tau_n^{\vert \widehat \rho_n-\rho\vert}-1\right),\\
\end{eqnarray*}
\noindent
where the first inequality comes from the fact that $\left|{e^x-1\over x}-1\right| \leq e^{|x|}-1.$ 
\noindent
Hence
\begin{eqnarray*}
{\sqrt {k_n} \over {\log ({n / k_n}) \log (\tau_n)}} \, B_{6,k_n} &=& {\sqrt {k_n} \over {\log ({n / k_n}) }} b\left(\log ({n / k_n})\right) \\
&\times&\left(\widehat \rho_n-\rho\right) \, {\int_1^{\tau_n} x^{\rho-1}\, \log (x) \, dx \over \log (\tau_n)} \left\{1+O\left(\tau_n^{|\widehat \rho_n-\rho|}-1\right)\right\}.
\end{eqnarray*}
\noindent
If $\tau_n \to \infty$, this implies that $\widehat \rho_n \stackrel{\mathbb P}{\rightarrow} \rho$ by the assumption (\ref{C3}) and therefore
\begin{eqnarray}
{\sqrt {k_n} \over {\log ({n / k_n}) \log (\tau_n)}} \, B_{6,k_n} = o_{\mathbb P}(1).
\label{terme6-1}
\end{eqnarray}
\noindent
Combining (\ref{terme1}), (\ref{terme2}) and (\ref{terme3}) with (\ref{terme4})-(\ref{terme6-1}), Theorem~\ref{thun}(i) follows.  
\noindent
If $\tau_n \to \tau, \tau>1$, then the normalization factor 
$\log(\tau_n)\to\log(\tau)\neq 0$ can be omitted in (\ref{terme1}), 
(\ref{terme3}) and (\ref{terme4}) while preserving the negligeability of these terms.
Besides, 
we can replace $\widehat \rho_n$ with
any canonical choice, for instance $\rho^\#<0$, and therefore
\begin{eqnarray}
\nonumber
\frac{\sqrt {k_n}}{\log ({n / k_n}) } \, B_{6,k_n} &=& {\sqrt {k_n} \over {\log ({n / k_n}) }} b\left(\log ({n / k_n})\right) \left(K_{\rho^\#}(\tau_n)-K_{\rho}(\tau_n)\right)\\
\label{terme6-2}
&\longrightarrow& \lambda\mu(\tau):=\lambda \left(K_{\rho^\#}(\tau)-K_{\rho}(\tau)\right).
\end{eqnarray} 
\noindent
The limiting distribution is then given by (\ref{terme2}) and (\ref{terme5}) with a bias term due to (\ref{terme6-2}). To conclude with the second part of our Theorem~\ref{thun}, we have to establish the limiting distribution of
\begin{eqnarray*}
U_n:={\sqrt {k_n} K_{\rho^\#}(\tau_n) \over \log({n/ k_n})} \left(\widehat b\left(\log({n/ k_n})\right)-b\left(\log({n/ k_n})\right)\right) +{\sqrt {k_n}\log(\tau_n) \over \log({n/ k_n})} \left(\widehat \theta_n-\theta\right).
\end{eqnarray*}
To this aim, remark that
$$U_n = {k_n^{-{1 \over 2}} \over \log ({n/ k_n})} \sum_{j=1}^{k_n}
\omega_{j,n} (f_j-1) + o_{\mathbb P}(1),$$
where
$$\omega_{j,n}=\beta_{j,n} K_{\rho^\#}(\tau_n) +\alpha_{j,n}\log(\tau_n)$$
and
$$\alpha_{j,n} = \Bigr(\theta+b\left(\log({n/ k_n})\right) x_j\Bigr)\left(1-{x_j-\overline x_{k_n} \over {1\over k_n} \sum_{i=1}^{k_n}(x_i-\overline x_{k_n})^2} \overline x_{k_n}\right).$$
Using Lemma~7.3 in Diebolt et al. (2006), direct computations lead to
\begin{eqnarray*}
\sum_{j=1}^{k_n} \mbox{Var}\left(\omega_{j,n}(f_j-1)\right)
=\sum_{j=1}^{k_n}\omega_{j,n}^2 \sim \theta^2 \left(\log({n/ k_n})\right)^2 \, k_n \, \sigma^2(\tau) 
\end{eqnarray*}
and
\begin{eqnarray*}
\sum_{j=1}^{k_n} \mathbb E \left(\omega_{j,n}(f_j-1)\right)^4 = 9 
\sum_{j=1}^{k_n} \omega_{j,n}^4 \sim C k_n \, \left(\log({n/ k_n})\right)^4,
\end{eqnarray*}
where $C$ is a suitable constant. Therefore a direct application of Lyapounov's theorem yields
$$U_n \stackrel{d}{\longrightarrow} {\cal N}\left(0, \theta^2 \sigma^2(\tau)\right),$$
which achieves the proof of the second part of Theorem~\ref{thun}. \hfill{\cqfd}\\

\noindent
{\bf Proof of Theorem~\ref{thdeux}.} We have:
\begin{eqnarray*}
 \ & \ & \log (\widetilde x_{p_n}) - \log(x_{p_n}) - b(\log(n/k_n)) \log(\tau_n)\frac{1}{k_n} \sum_{j=1}^{k_n} \left ( \frac{\log(n/j)}{\log(n/k_n)} \right )^{\rho} + b(\log(n/k_n)) \frac{\tau_n^{\rho}-1}{\rho} \\
 \ & = & \theta \{\log ( - \log(U_{k_n,n}))- \log \log (n/k_n)\} \\
 \ & + & \left ({\widetilde{\theta}}_n - \theta - b(\log n/k_n) \frac{1}{k_n} \sum_{j=1}^{k_n} \left ( \frac{\log(n/j)}{\log(n/k_n)} \right )^{\rho} \right ) \log(\tau_n) \\
 \ & + & \{ \log \ell( -\log(U_{k_n,n})) - \log \ell (\log(n/k_n)) \} \\
 \ & + & \left \{ \log \ell (\log(n/k_n)) - \log \ell (-\log(p_n)) + b(\log(n/k_n)) \frac{\tau_n^{\rho}-1}{\rho} \right \} \\
 \ & := & B_{1,k_n} + B_{7,k_n} + B_{3,k_n} + B_{4,k_n}.
\end{eqnarray*}
From (\ref{terme1}), we have
\[ \frac{\sqrt{k_n}}{\log(\tau_n)} B_{1,k_n} = O_{\rm{P}} \left ( \frac{1}{\log(n/k_n)\log(\tau_n)} \right ) = o_{\rm{P}}(1), \]
and Theorem~2.2 in Diebolt et al. (2006) states that
\[ \frac{\sqrt{k_n}}{\log(\tau_n)} B_{7,k_n}  \stackrel{d}{\longrightarrow}  {\cal{N}}(0,\theta^2). \]
From (\ref{terme3-0}), we have
\[ B_{3,k_n} = O_{\rm{P}} \left ( \frac{b(\log(n/k_n))}{\sqrt{k_n} \log(n/k_n)} \right ), \]
and thus $\sqrt{k_n}b(\log(n/k_n)) \to \lambda\in{\mathbb{R}}$ entails
\[ \frac{\sqrt{k_n}}{\log(\tau_n)} B_{3,k_n} = O_{\rm{P}} \left ( \frac{1}{\sqrt{k_n} \log(n/k_n)\log(\tau_n)} \right ) = o_{\rm{P}}(1). \]
Finally, (\ref{terme4-0}) implies that
\[ B_{4,k_n} = o_{\rm{P}}(b(\log(n/k_n))), \]
which, combined with  $\sqrt{k_n}b(\log(n/k_n)) \to \lambda\in{\mathbb{R}}$ yields
\[ \frac{\sqrt{k_n}}{\log(\tau_n)} B_{4,k_n} = o_{\rm{P}}\left ( \frac{1}{\log(\tau_n)} \right ) = o_{\rm{P}}(1), \]
which achieves the proof of Theorem~\ref{thdeux}. \hfill{\cqfd}\\ 

\section*{Appendix}

Diebolt et al. (2006) introduced
the class of distributions ${\cal D}(\alpha,\beta)$
with distribution function given by
$$
1-F(x) = \exp(-H(x))  \mbox{ where }  H^{-1}(x):= 
x^{1/\alpha} (1+x^{-\beta}),
$$
$\alpha$ and $\beta$ being two parameters such that
$0<\alpha$, $0< \beta< 1$ and $\alpha\beta\leq 1$.
Under these conditions, the above
class of distributions fulfill assumptions
(\ref{model}) with $(R_{\ell}(b, \rho))$ and (\ref{hypob})
where $\theta=1/\alpha$, $\rho=-\beta$,
$\ell(x)=1+x^{-\beta}$ and
$b(x)=-\beta x^{-\beta}$.  It is thus possible
to obtain distributions with arbitrary $\theta>0$
and $-1<\rho<0$.   These results are
summarized in Table~\ref{tabex}.

\section*{Acknowledgement}

The authors are grateful to the referees for a careful reading
of the paper that led to significant improvements of the earlier
draft.  



\begin{table}[h]
\begin{center}
$
\begin{array}{|c|c|c|c|}
\hline
&&&\\
\mbox{Distribution}                       & \theta & b(x) & \rho \\
&&&\\
\hline
&&&\\
\mbox{Absolute Gaussian }|{\mathcal N}|(\mu,\sigma^2)      & 1/2   & \displaystyle\frac{1}{4} \frac{\log x}{x} & -1 \\
&&&\\
\mbox{Gamma }\Gamma(\alpha\neq 1,\beta)      & 1     & (1-\alpha) \displaystyle \frac{\log x}{x} & -1 \\
&&&\\
\mbox{Weibull }{\mathcal W}(\alpha,\lambda)    & 1/\alpha & 0 & -\infty \\
&&&\\
{\mathcal D}(\alpha,\beta)    & 1/\alpha & -\beta x^{-\beta} & -\beta \\
&&&\\
\hline
\end{array}
$
\end{center}
\caption{Parameters $\theta$, $\rho$ and the function $b(x)$ associated
to some  distributions}
\label{tabex}
\end{table}

\clearpage 





\end{document}